\documentclass[11pt,fleqn]{article} 
\usepackage{graphicx}
\usepackage{textcomp} 
\usepackage{epstopdf}
\usepackage{latexsym}
\usepackage{amsmath}
\usepackage{amssymb}
\usepackage{bm}
\usepackage[mathcal]{euscript}
\usepackage{slashed}

\usepackage[top=30mm, bottom=30mm, left=25mm, right=25mm]{geometry}

\numberwithin{equation}{section}

\usepackage{marvosym}

\usepackage{indentfirst}

\newcommand\blfootnote[1]{
  \begingroup
  \renewcommand\thefootnote{}\footnote{#1}
  \addtocounter{footnote}{-1}
  \endgroup
}

\usepackage{hyperref}
\hypersetup{
        unicode,	
        colorlinks,
        citecolor=blue,
        linkcolor=blue, 
        urlcolor=red,
        bookmarksopen=true,
        bookmarksopenlevel=\maxdimen,
      }

\def\gl#1#2{\ifmmode \mathrm{GL}(#1; {\bf #2}) \else $\mathrm{GL}(#1; {\bf #2})$\fi}
\def\sl#1#2{\ifmmode \mathrm{SL}(#1; {\bf #2}) \else $\mathrm{SL}(#1; {\bf #2})$\fi}
\def\so#1{\ifmmode \mathrm{SO}({#1}) \else $\mathrm{SO}(#1)$\fi}

\def\sp#1#2{\ifmmode \mathrm{Sp}(#1; {\bf #2}) \else $\mathrm{Sp}(#1; {\bf #2})$\fi}
\def\usp#1{\ifmmode \mathrm{USp}(#1) \else $\mathrm{USp}(#1)$\fi}
\def\spin#1{\ifmmode \mathrm{Spin}(#1) \else $\mathrm{Spin}(#1)$\fi}
\def\su#1{\ifmmode \mathrm{SU}({#1}) \else $\mathrm{SU}(#1)$\fi}

\def\double #1{#1{\hbox{\kern-2pt $#1$}}}

\mathcode`\*="702A                  
\def\half{{\textstyle{1\over{\raise.1ex\hbox{$\scriptstyle{2}$}}}}}

\def \p{\partial}

\def \a{\alpha}

\def \b{\beta}

\def \e{\epsilon}
\def \d{\delta}

\def \g{\gamma}

\def \l{\lambda}

\def \o{\omega}

\def \O{\Omega}
\def\Oh{\widehat\Omega}
\def \t{\theta}

\def \N{\nabla}

\def\Nh{\widehat\nabla}

\def \r{\rho}

\def\s{\sigma}

\def\Rh{\widehat R}
\def\Th{\widehat T}

\begin{document}

\begin{flushright}
\makebox[0pt][b]{}
\end{flushright}

\vspace{40pt}
\begin{center}
{\LARGE Compactifications of Type II Supergravities in Superspace}

\vspace{40pt}
Osvaldo Chandia${}^{\clubsuit}$ and Brenno Carlini Vallilo${}^{\spadesuit}$
\vspace{40pt}

{\em 
${}^{\clubsuit}$ Departamento de Ciencias, Facultad de Artes Liberales \\ Universidad Adolfo Ib\'a\~nez, Diagonal Las Torres 2640, Pe\~nalol\'en, Chile }\\

\vspace{20pt}

{\em 
${}^{\spadesuit}$ Departamento de Ciencias F\'{\i}sicas, Facultad de Ciencias Exactas\\ Universidad Andr\'es Bello, Sazi\'e 2212, Santiago, Chile}

\vspace{60pt}
{\bf Abstract}
\end{center}
We propose a way to describe compactifications of Type II supergravities with fluxes directly from superspace. The superspace used is the one that arises naturally from the pure spinor superstring. We show how previous results of flux compactifications can be obtained from our method.  

\blfootnote{
${}^{\clubsuit}$ \href{mailto:ochandiaq@gmail.com}{ochandiaq@gmail.com},    
${}^{\spadesuit}$ \href{mailto:vallilo@unab.cl}{vallilo@unab.cl} }

\setcounter{page}0
\thispagestyle{empty}

\newpage

\tableofcontents

\parskip = 0.1in
\section{Introduction and motivation}
\label{intro}
Supersymmetry is pivotal in superstring theory, underlying many of its most intriguing properties. Supersymmetry enables exact computations and nontrivial checks of the web of dualities connecting different string theories. Just as space-time provides the natural geometric framework for formulating Lorentz and generally covariant theories, superspace is the ideal mathematical setting for describing supersymmetric field theories and supergravity models. The additional fermionic dimensions of superspace elegantly incorporate supersymmetric transformations. As such, superspace would be an invaluable tool for constructing, analyzing, and connecting supersymmetric string vacua. Much as space-time geometry provided vital insights leading to general relativity, the geometry of superspace has the potential to furnish critical clues needed for a better understanding of string theory. 

Nonetheless, despite its long history, the use of superspace in string theory has been minimal. One technical difficulty is that spinor representations have different properties for different space-time dimensions. Related to that, the superspace description of supersymmetric multiplets varies widely depending on the number of supersymmetries. Perhaps the most critical obstacle is obtaining an off-shell formulation for theories with sixteen or more supersymmetries.   

The use of superspace becomes challenging to avoid when considering Ramond-Ramond backgrounds. In the RNS formulation, they couple to the world-sheet vertex operators containing spin fields, which are operators that change the RNS fermions' boundary conditions and also have non-zero picture numbers. These two properties render a world-sheet description of finite RR backgrounds rather strenuous \cite{Berenstein:1999jq, Berenstein:1999ip}. One way to partially solve these problems comes in the form of the so-called hybrid formalism \cite{Berkovits:1994wr}. Suppose one is interested in only maintaining Poincar\'e covariance in lower dimensions. In that case, it is possible to choose the world-sheet super-charges such that their anti-commutators give the translation operator in the zero picture. After some field redefinitions, the world-sheet variables will transform linearly under supersymmetries. The hybrid formalism was used in the description of RR background in {\it e.g.}, \cite{Berkovits:1999im, Berkovits:1999xv, Berkovits:2001tg, Linch:2006ig, Linch:2008rw}. A promising way to describe RR backgrounds using the RNS formalism is superstring field theory (SSFT) \cite{Sen:2014pia, Sen:2014dqa,Sen:2015hha,deLacroix:2017lif}. This approach represents the background as a classical solution of the SSFT equations of motion. As far as we know, the first paper that explored this idea for Ramond-Ramond backgrounds was \cite{Cho:2018nfn}. More recently, a very detailed work describing a case of flux compactification appeared \cite{Cho:2023mhw}. A different approach to studying flux backgrounds using superspace was discussed in \cite{Grassi:2006cd}.

In this work, we study compactifications of Type II supergravity theories that preserve $N=1$ supersymmetry in four dimensions working directly from superspace. The literature on the subject is too vast to give a proper list of references. The reviews \cite{Grana:2005jc, Koerber:2010bx} provide an excellent (although not up-to-date) summary of the subject and contain an extensive list of references. Nevertheless, we will cite works directly related to some of our conclusions in the body of the paper. The curved superspace formulation we use comes from the pure spinor superstring \cite{Berkovits:2001ue}. This formulation is non-standard; it has two connections and a scaling symmetry that acts only on the spinors. Of course, it can be mapped to standard descriptions, but since we want to apply the results to the pure spinor superstring, we will use the more adequate formalism.  

The usual way to study\footnote{Our notation and more precise definitions, together with the proof of some claims, will be given in the next section.} symmetries in a curved (super)space to introduce a Killing supervector field 
\begin{align} 
    K= \xi^A \nabla_A +\frac{1}{2} \Lambda^{ab} M_{ab} + \sigma S +\widetilde\sigma\widetilde S,
\end{align}
where $M_{ab}$ are the Lorentz generators and $(S,\widetilde S)$ are scale generators that act only on spinors. The defining condition the Killing supervector satisfies is 
\begin{align}\label{Killing}
    [K,\nabla_A]=0. 
\end{align}
If the theory is purely geometrical, this is sufficient. However, ten-dimensional supergravity theories have additional matter fields. Another characteristic of these theories is that they have a scalar superfield $\Phi$ whose first component is the dilaton and that almost all gauge covariant tensors (the only exception being the Ramond-Ramond five form in Type IIB) can be found at higher $\theta$ components. Therefore, for a general supergravity background to be invariant under the transformation generated by $K$, it is also necessary to require 
\begin{align}\label{DeltaDilaton}
    \delta\Phi=K(\Phi)=0.
\end{align}

In a previous work \cite{Chandia:2022uyy}\footnote{See also \cite{Chandia:2009it, Chandia:2011wd} for earlier works.}, we studied heterotic string backgrounds with $N=1$ supersymmetry in four dimensions directly from the ten-dimensional superspace using (\ref{Killing}) and (\ref{DeltaDilaton}). Usually, given a fixed background, the Killing supervectors are found from (\ref{Killing}). Instead, we wanted to find the conditions on superspaces if we demand that the Killing supervectors have some desired form. For example, for the space to have translation invariance in some of its dimensions, we should require that 
$\xi^a|_{\theta=0}\neq 0$ for and $\nabla_A \xi^a|_{\theta=0}=0$ $a=0,1,2,3$. Then, the higher $\theta$ components of $\xi^A$ can be found solving (\ref{Killing}) in terms of the general background. This process will also fix to zero the $|_\theta=0$ projection of some background tensors. The same steps can be applied to Type II supergravities. 

Although this process will help understand the curved superspace of compactifications, it will not help study the first important set of observables: the massless physical states. In the pure spinor formalism the physical states are described by operators with ghost number $2$ in the cohomology of the BRST operator $Q=\int d\sigma(\lambda^\alpha d_\alpha + \widetilde\lambda^{\widetilde\alpha}\widetilde d_{\widetilde\alpha})$, where $(\lambda,\widetilde\lambda)$ are the pure spinor ghosts and $(d_\alpha,\widetilde d_{\widetilde\alpha})$ are the conjugate momenta for the odd superspace variables $(\theta,\widetilde\theta)$. At the lowest order in $\alpha'$, the action of $Q$ can be described by the differential operator 
\begin{align}\label{brstdifferential}
    Q= \lambda^\alpha \nabla_\alpha + \widetilde\lambda^{\widetilde\alpha} \widetilde\nabla_{\widetilde\alpha}.
\end{align}
where $(\nabla_\alpha, \widetilde\nabla_{\widetilde\alpha})$ are the odd superspace covariant derivatives. The massless vertex operators can be written as 
\begin{align}
    {\mathcal V}= \lambda^\alpha \widetilde\lambda^{\widetilde\alpha} A_{\alpha\widetilde\alpha}(Z),
\end{align}
where $A_{\alpha\widetilde\alpha}(Z)$ is a tensor superfield satisfying the equations coming from $Q{\mathcal V}=0$ and the gauge transformations from $\delta {\mathcal V}= Q\xi$, for some appropriate $\xi$. In the case of a compactification preserving $N=1$ supersymmetry in four dimensions, the fields describing the fluctuations of the internal geometry are organized into four-dimensional chiral and anti-chiral multiplets. It is natural to ask if describing these multiplets in a standard four-dimensional superfield is possible. In other words, the vertex operator should factorize\footnote{We will use the $\mathfrak{sl}(2,\mathbb{C})+\mathfrak{su}(4)$ decomposition for the ten-dimensional spinors. For a longer explanation of the notation used in this paper, see the end of the present section \ref{Notation}.} 
\begin{align}\label{factorizedVertex}
    {\mathcal V} = \sum_{\Gamma}\left(\Phi^\Gamma_{\alpha\widetilde\alpha}(Z_{d=4}) \Psi^{\Gamma}_{I\widetilde I}(Y)\lambda^{\alpha I} \widetilde\lambda^{\widetilde{\alpha I}}+{\rm c.c.}\right),
\end{align}
where $\Gamma$ is a generic index for the different compactification moduli, $\Psi^\Gamma_{I\widetilde I}(Y)$ is some internal ``superspace harmonic form'' and $\Phi^\Gamma_{\alpha\widetilde\alpha}(Z_{d=4})$ is the desired four dimensional superfield. The equations of motion for the fields $\Phi^\Gamma$ should also have the standard superspace form. For that to be possible, we must describe how the usual four-dimensional odd covariant derivatives appear inside $Q$. In the $\mathfrak{sl}(2,\mathbb{C})+\mathfrak{su}(4)$ notation, the BRST differential (\ref{brstdifferential}) is written as 
\begin{align}
    Q=\lambda^{\alpha I}\nabla_{\alpha I}+\lambda^{\dot\alpha}{}_I\nabla_{\dot\alpha}{}^I+ \widetilde\lambda^{\alpha\widetilde I}\widetilde\nabla_{\alpha\widetilde I}+\widetilde\lambda^{\dot\alpha}{}_{\widetilde I}\widetilde\nabla_{\dot\alpha}{}^{\widetilde I}.
\end{align}
The position of the index $\widetilde I$ depends on whether we choose type IIA or IIB. The first step for our goal is to introduce a pair of complex bosonic $SU(4)$ spinors $(\chi^I,\widetilde\chi^{\widetilde I})$ superfields and define the following covariant derivatives 
\begin{align}\label{newDerivatives}
    &\nabla_\alpha = \chi^I\nabla_{\alpha I}+\widetilde\chi^{\widetilde I}\widetilde\nabla_{\alpha \widetilde I},\\ &\Delta_\alpha=\chi^I\nabla_{\alpha I}-\widetilde\chi^{\widetilde I}\widetilde\nabla_{\alpha \widetilde I},\quad
    \Delta_{\alpha k} = \chi_I \sigma_{ k}^{IJ}\nabla_{\alpha J},\quad\widetilde\Delta_{\alpha k}=\widetilde\chi_{\widetilde I}\widetilde\sigma_{k}^{\widetilde I\widetilde J}\widetilde\nabla_{\alpha \widetilde J},\label{mixed_derivatives}
\end{align}
and their complex conjugates. The full covariant derivatives can be recovered as 
\begin{align}
    &\nabla_{\alpha I}= \frac{1}{2|\chi|^2}\left( \chi_I(\nabla_\alpha +\Delta_\alpha) -\sigma^{k}_{IJ}\chi^J\Delta_{\alpha{k}}\right),\\
    &\widetilde\nabla_{\alpha\widetilde I}= \frac{1}{2|\widetilde\chi|^2}\left( \widetilde\chi_I(\nabla_\alpha -\Delta_\alpha) -\widetilde\sigma^k_{\widetilde I\widetilde J}\widetilde\chi^{\widetilde J}\widetilde\Delta_{\alpha k} \right).
\end{align}
Which can be checked using the the identities $\sigma_k^{IJ}\sigma^k_{KL}=-2(\d^I_K\d^J_L-\d^I_L\d^J_K)$ and its analog for $\widetilde\sigma^\mathsf{k}$. The set (\ref{newDerivatives}) has sixteen complex linearly independent derivatives with the choices $\sigma^{\bar{\mathsf{k}}}_{IJ}\chi^J=\widetilde\sigma^{\bar{\mathsf{k}}}_{\widetilde I\widetilde J}\widetilde\chi^{\widetilde J}=0$. 

We are going to define $(\nabla_\alpha,\nabla_{\dot\alpha})$ as the ``four-dimensional'' covariant derivatives and 
$\allowbreak (\Delta_\alpha,\allowbreak \Delta_{\dot\alpha},\allowbreak \Delta_{\alpha k},\allowbreak \Delta_{\dot\alpha k },\allowbreak \widetilde\Delta_{\alpha k},\widetilde\Delta_{\dot\alpha k })$ as the ``mixed'' covariant derivatives. This distinction is because it is impossible to eliminate the four-dimensional derivative on the left-hand side of the anti-commutators of the mixed covariant derivatives. 
In this approach, an ``internal'' superfield will be defined as being annihilated by all ``four-dimensional'' derivatives and Lorentz generators. 

If we define 
\begin{align}
&\lambda^\alpha= \frac{1}{2|\chi|^2} \left(\chi_I\lambda^{\alpha I}+\widetilde\chi_{\widetilde I}\widetilde\lambda^{\alpha\widetilde I}\right), \quad 
\Lambda^{\alpha}=\frac{1}{2|\widetilde\chi|^2} \left(\chi_I\lambda^{\alpha I}-\widetilde\chi_{\widetilde I}\widetilde\lambda^{\alpha\widetilde I}\right),\\
&\Lambda^\alpha_{{{k}}}= -\frac{1}{2|\chi|^2}(\sigma_{{{k}}})_{KL}\chi^K\lambda^{\alpha L},\quad 
\widetilde\Lambda^\alpha_{{{k}}}= -\frac{1}{2|\widetilde\chi|^2}(\sigma_{{{k}}})_{\widetilde K\widetilde L}\widetilde\chi^{\widetilde K}\widetilde\lambda^{\alpha \widetilde L},
\end{align}
we can then write the BRST charge as 
\begin{align}
    Q=\lambda^\alpha\nabla_\alpha +\Lambda^\alpha\Delta_\alpha+\Lambda^\alpha_k\nabla_{\alpha {k}}+\widetilde\Lambda^\alpha_k\widetilde\nabla_{\alpha k} + {\rm c.c.},
\end{align}

The idea is to impose that the four-dimensional super-covariant derivatives generate the covariant derivative algebra of either flat space of $AdS_4$. $(\nabla_\alpha,\nabla_{\dot\alpha})$ are not the most general possibility; other terms could be added. However, we will find that $(\chi^I,\widetilde\chi^{\widetilde I})$ superfields provide conditions that are general enough.  

This work is organized as follows. In Section \ref{PureSugraReview}, we review ten dimensional Type II supergravities in the pure spinor formulation. We describe how various tensors are determined by the Berkovits-Howe constraints. In Section \ref{4Dsusy}, we derive the conditions imposed on the curved supergeometry by requiring $N=1$ four-dimensional supersymmetry in $AdS_4$ or flat space. 
In Section \ref{possibilities}, we discuss how the conditions derived in Section \ref{4Dsusy} can be solved with a nontrivial compactification. We conclude the paper and discuss future work in Section \ref{conclusion}.

\paragraph{\textbf{A warning about notation}}\label{Notation} This work deals with various types of indices. We decided to use different conventions in different sections to reduce the number of index decorations. In Section \ref{PureSugraReview}, there are only ten-dimensional tensors and spinors; lowercase Latin letters at the beginning of the alphabet $a, b, c,\cdots$ are ten-dimensional local Lorentz indices. Lowercase Greek letters at the beginning of the alphabet $\alpha, \beta, \gamma, \cdots$ are ten-dimensional spinor indices from $1$ to $16$. The meaning of the spinor indices with a tilde depends on whether it is Type IIA or Type IIB. Tildes will also differentiate left and right-moving sectors in the world-sheet variables {\it, e.g.} $(\partial,\widetilde\partial)$. In Sections \ref{4Dsusy} and \ref{possibilities}, we need indices to describe the four-dimensional superspace and indices to describe the internal six-dimensional space. In these sections, we use lowercase Latin letters at the beginning of the alphabet $a, b, c,\cdots$ for four-dimensional local Lorentz indices. Furthermore, undotted and dotted lowercase Greek letters at the beginning of the alphabet $\alpha,\beta,\gamma,\cdots$ are four-dimensional chiral and anti-chiral spinor indices from $1$ to $2$; uppercase Latin letter in the middle of the alphabet $I, J, K,\cdots$ are six-dimensional spinor indices from $1$ to $4$. Real tangent space indices for the internal space will be 
$i, j, k, \cdots$ from $1$ to $6$, and complex internal indices will be unbarred and barred lowercase Latin letters in upright San Serif font $\mathsf{k}, \mathsf{l}, \mathsf{m}, \cdots$ from $1$ to $3$. Finally, when referring to the full ten-dimensional set of tangent space-time indices is necessary, we will use an underline as $\underline{a}$. Some useful identities are listed in Appendix \ref{identities}.

\section{Ten-dimensional Type II supergravities form the pure spinor string}
\label{PureSugraReview}
Consider the type II superspace in ten dimensions used in the pure spinor formalism. The geometry of this superspace is constrained by the pure spinor BRST symmetry, which is generated by the charges
\begin{align}\label{Qbrst}
    Q=\oint d\s \l^\a d_\a,\quad {\widetilde Q}=\oint d\s {\widetilde\l}^{\widetilde\a}{\widetilde d}_{\widetilde\a},
\end{align}
where $(\l,{\widetilde\l})$ are the pure spinor variables and $(d,{\widetilde d})$ are the world-sheet generators of translations in superspace. The BRST invariant world-sheet action is given by
\begin{align}\label{SII}
    S=\int d^2\s &\left(\frac12 \Pi_a\widetilde\Pi^a+\frac12\Pi^A\widetilde\Pi^B B_{BA}+d_\a\widetilde\Pi^\a+\widetilde{d}_{\widetilde\a}\Pi^{\widetilde\a}+\o_{\a}{\widetilde \partial}\l^\a+{\widetilde\o}_{\widetilde\a}\partial{\widetilde\l}^{\widetilde\a}\right.\cr
    &\left.  
    +\l^\b\widetilde\p Z^M \O_{M\b}{}^\a\o_{\a}+\widetilde\lambda^{\widetilde\b} \p Z^M \Oh_{M\widetilde\b}{}^{\widetilde\a}{\widetilde\o}_{\widetilde\a}+d_\a{\widetilde d}_{\widetilde\b}P^{\a\widetilde\b}\right.\cr &\left.+\l^\a\o_\b{\widetilde d}_{\widetilde\g} C_\a{}^{\b\widetilde\g}+{\widetilde\l}^{\widetilde\a}{\widetilde\o}_{\widetilde\b}d_\g C_{\widetilde\a}{}^{\widetilde\b\g}+\l^\a\o_\b{\widetilde\l}^{\widetilde
    \g}{\widetilde\o}_{\widetilde\r}S_{\a\widetilde\g}{}^{\b\widetilde\r}\right),
\end{align}
where the background fields are functions of the superspace coordinates are $Z^M=(X^m, \t^\mu, \widetilde\theta^{\widetilde\mu})$ where $m=0,\dots 9\; ; \mu,\widetilde\mu=1,\dots 16$ are curved superspace indices. The fermionic indices $\mu$ and $\widetilde\mu$ have the opposite chirality for the type IIA case and have the same chirality for type IIB case. This means that $\widetilde\theta^{\widetilde\mu}=\widetilde\theta_\mu$ for the type IIA case and $\widetilde\theta^{\widetilde\mu}=\widetilde\theta^\mu$ for the type IIB case. The super-coordinates enter in the world-sheet action in the combinations $\Pi^A=\p Z^M E_M{}^A$ and $\widetilde\Pi^A=\widetilde\p Z^M E_M{}^A$ where $\p, \widetilde\p$ are the derivatives respect to the world-sheet coordinates and $E_M{}^A$ is the super-vielbein and rotates the super-space index $M$ into the local super-space index $A$. This local index ranges in $(a, \a, \widetilde\a)$. The world-sheet action includes the covariant derivatives of the pure spinor variables $\l^\a$ and $\widetilde\lambda^{\widetilde\a}$, they are
\begin{align}
{\widetilde D}\l^\a=\widetilde\p\l^\a+\l^\b\widetilde\p Z^M \O_{M\b}{}^\a,\quad D\widetilde\lambda^{\widetilde\a} = \p\widetilde\lambda^{\widetilde\a}+\widetilde\lambda^{\widetilde\b} \p Z^M \Oh_{M\widetilde\b}{}^{\widetilde\a} ,
\label{1}
\end{align}
where $\O_{M\a}{}^\b$ and $\Oh_{M\widetilde\b}{}^{\widetilde\a}$ are the background connections. To respect a symmetry involving the conjugate variable of $\l$ and the conjugate variable of $\widetilde\lambda$, these connections have the form
\begin{align}
\O_{M\a}{}^\b =\d_\a^\b \O_M + \frac14(\g^{ab})_\a{}^\b \O_{Mab},\quad \Oh_{M\widetilde\a}{}^{\widetilde\b} =\d_{\widetilde\a}^{\widetilde\b} \Oh_M + \frac14(\g^{ab})_{\widetilde\a}{}^{\widetilde\b} \Oh_{Mab},
\label{DefConnections}
\end{align}
where $\g^{ab}$ is the antisymmetric product of two symmetric $16\times16$ $\g$-matrices.
Usually, there is only one connection, the Lorentz connection $\O_{Mab}$. It turns out that $\O_M$ and $\Oh_M$ are related to the background dilaton superfield $\Phi$ and $\Oh_{Mab}$ is a function of $\O_{Mab}$ and torsion components, as we show below. This reduction only happens after gauge fixing 
\cite{Berkovits:2001ue} the three independent local Lorentz symmetries that the action (\ref{SII}) suffers. These five local gauge symmetries generate some apparent puzzles; the fact that there are two different curvatures and that gamma matrices are not invariant tensors. This last fact implies that the covariant derivatives (defined with the connections directly from (\ref{SII})) of the gamma matrices do not vanish \cite{Chandia:2022uyy}. 

The nilpotency of the BRST charges (\ref{Qbrst}) together with the holomorphicity of $\l^\a d_\a$ and the anti-holomorphicity of ${\widetilde\l}^{\widetilde\a}{\widetilde d}_{\widetilde\a}$ put the background fields to satisfy the constraints of type II supergravity in ten dimensions \cite{Berkovits:2001ue}. These constraints are expressed in terms of torsion and curvature components. As usual, we define the super one forms $E^A=dZ^M E_M{}^A$, $\O_\a{}^\b=dZ^M \O_{M\a}{}^\b$,  $\Oh_{\widetilde\a}{}^{\widetilde\b}=dZ^M \Oh_{M\widetilde\a}{}^{\widetilde\b}$. From this, we define the curvature and torsion 2-forms. The curvatures are
\begin{align}
R_\a{}^\b=d\O_\a{}^\b + \O_\a{}^\g \O_\g{}^\a,\quad \Rh_{\widetilde\a}{}^{\widetilde\b}=d\Oh_{\widetilde\a}{}^{\widetilde\b}+\Oh_{\widetilde\a}{}^{\widetilde\g} \Oh_{\widetilde\g}{}^{\widetilde\b} ,
\label{3}
\end{align}
here the wedge product is assumed. The torsion is the covariant derivative of the vielbein one-form, for $T^\a$ and $T^{\widetilde\a}$ we have
\begin{align}
T^\a=\N E^\a = d E^\a + E^\b \O_\b{}^\a,\quad T^{\widetilde\a}=\N E^{\widetilde\a}=d E^{\widetilde\a}+E^{\widetilde\b}\Oh_{\widetilde\b}{}^{\widetilde\a} ,
\label{4}
\end{align}
where, again, the wedge product is assumed. There are two possible definitions of torsion as the covariant derivative of $E^a$. The first choice uses $\O^{ab}$ as the connection, and the second uses $\Oh^{ab}$ as a connection. Then we have
\begin{align}
T^a=\N E^a=d E^a + E^b \O_b{}^a,\quad \Th^a=\Nh E^a=d E^a + E^b \Oh_b{}^a .
\label{5}
\end{align}
Note that $\Th_{AB}{}^c=T_{AB}{}^c$ unless $A$ or $B$ takes the value $a$ or $b$. Similarly, two possible curvatures are defined with $\O^{ab}$ or $\Oh^{ab}$. They are
\begin{align}
R_a{}^b=d\O_a{}^b+\O_a{}^c\O_c{}^b,\quad \Rh_a{}^b=d\Oh_a{}^b+\Oh_a{}^c\Oh_c{}^b .
\label{6}
\end{align}
The torsions of  (\ref{3}) satisfy the Bianchi identities, they are
 \begin{align}
\N T^\a=E^\b R_\b{}^\a,\quad \N T^{\widetilde\a}=E^{\widetilde\b} \Rh_{\widetilde\b}{}^{\widetilde\a} ,
\label{7}
\end{align}
and the torsions in (\ref{5}) satisfy the identities
\begin{align}
\N T^a=E^b R_b{}^a,\quad \Nh \Th^a=E^b \Rh_b{}^a .
\label{8}
\end{align}
From the superspace point of view, the curvature $\widehat R_a{}^b$ and torsion $\widehat T^a$ are superfluous; they do not contain any new geometric information. Below, we will show how they relate to other torsion and curvature components. Some constraints to have an on-shell superspace are simpler when written in terms of $\widehat R_a{}^b$ and $\widehat T^a$. Furthermore, they are also essential since some sigma model couplings are given by these tensors in a background field expansion \cite{Chandia:2003hn} \cite{Bedoya:2006ic}. Another background field is the Kalb-Ramond two-form $B$. Its exterior derivative is $H=dB$, and the Bianchi identity is $dH=0$. 

BRST symmetry imposes constraints on background torsion and $H$-components. We have $T_{AB}{}^C=H_{ABC}=0$ for $(A, B, C) \in (\a ,\widetilde\a)$ and
\begin{align}\label{gamma}
&T_{\a\widetilde\b}{}^a=H_{a\a\widetilde\a}=H_{ab\a}=H_{ab\widetilde\a}=T_{a\a}{}^\b=T_{a\widetilde\a}{}^{\widetilde\b}=0 \cr 
&T_{\a\b}{}^a=-\g^a_{\a\b},\quad T_{\widetilde\a\widetilde\b}{}^a=-\g^a_{\widetilde\a\widetilde\b},\quad H_{a\a\b}=-(\g_a)_{\a\b},\quad H_{a\widetilde\a\widetilde\b}=(\g_a)_{\widetilde\a\widetilde\b} .
\end{align}

We now use these constraints in Bianchi identities to obtain constraints on other torsion components. As it was done in \cite{Berkovits:2001ue}, Bianchi identities imply
\begin{align}
    &T_{\a ab}=2(\g_{ab}\O)_\a,\quad \Th_{\widetilde\a ab}=2(\g_{ab}\Oh)_{\widetilde\a},\cr 
    &\O_{\widetilde\a}=\Oh_\a=\O_a=\Oh_a=T_{\widetilde\a ab}=\Th_{\a ab}=0.
\end{align}

Consider the Bianchi identity $dH=0$. We have two options: they are
\begin{align}
\N_{[A} H_{BCD]}+\frac32 T_{[AB}{}^E H_{E CD] }= 0.
\label{H1}
\end{align}
or
\begin{align}
\Nh_{[A} H_{BCD]}+\frac32 \Th_{[AB}{}^E H_{E CD] }= 0.
\label{H2}
\end{align}
which are equivalent. These identities with $(A, B, C, D) \in (\a, \widetilde\a)$ and $(A, B, C, D) \in (a, \a, \widetilde\a)$ are trivially satisfied. Consider the identity (\ref{H1}) with  $(A, B, C, D) \in (a, b, \a, \b)$, it implies that $T_{abc}+H_{abc}=0$.  Similarly, the identity (\ref{H2}) with  $(A, B, C, D) \in (a, b, \widetilde\a, \widetilde\b)$, it implies that $\Th_{abc}-H_{abc}=0$.  At this point $\Oh_a{}^b$ is given by $\O_a{}^b$ and torsion components. In fact, 
\begin{align}
\Oh_{ca}{}^b = \O_{ca}{}^b - T_{ca}{}^b ,\quad \Oh_{\a a}{}^b = \O_{\a a}{}^b - T_{\a a}{}^b  ,\quad \Oh_{\widetilde\a a}{}^b = \O_{\widetilde\a a}{}^b + \Th_{\widetilde\a a}{}^b .
\label{OhOT}
\end{align}

We now add the remaining constraints for the background superfields. They are 
\begin{align}
&R_{\widetilde\a\widetilde\b\a}{}^\b=\Rh_{\a\b\widetilde\a}{}^{\widetilde\b}=0,\quad T_{a\a}{}^{\widetilde\a}=(\g_a)_{\a\b}{} P^{\b\widetilde\a},\quad T_{a\widetilde\a}{}^\a = -(\g_a)_{\widetilde\a\widetilde\b} P^{\a\widetilde\b} ,\cr
&C_\a{}^{\b\widetilde\a} = -\N_\a P^{\b\widetilde\a},\quad C_{\widetilde\a}{}^{\widetilde\b\a}=\N_{\widetilde\a} P^{\a\widetilde\b} ,\cr
&R_{a\widetilde\a\a}{}^\b=-(\g_a)_{\widetilde\a\widetilde\b} C_\a{}^{\b\widetilde\b},\quad \Rh_{a\a\widetilde\a}{}^{\widetilde\b}=-(\g_a)_{\a\b} C_{\widetilde\a}{}^{\widetilde\b\a} ,\cr
&S_{\a\widetilde\a}{}^{\b\widetilde\b}=\N_\a  C_{\widetilde\a}{}^{\b\widetilde\b} + \Rh_{\a\widetilde\g\widetilde\a}{}^{\widetilde\b} P^{\b\widetilde\g} = \N_{\widetilde\a} C_\a{}^{\b\widetilde\b} - R_{\widetilde\a\g\a}{}^\b P^{\g\widetilde\b} ,
\label{BHp}
\end{align}
where $P$ is the RR field-strength. 

As it was shown in \cite{Berkovits:2001ue} (see also \cite{Bedoya:2006ic}), 
\begin{align}\label{OmPhi}
    \O_\a=4\N_\a\Phi,\quad \Oh_{\widetilde\a}=4\N_{\widetilde\a}\Phi,
\end{align}
where $\Phi$ is the dilaton superfield.
Even derivatives of $\O_\a$ and $\Oh_{\widetilde\a}$ are connected to torsion components according to \cite{Bedoya:2006ic}
\begin{align}
\g^b_{\a\b}{} T_{ab}{}^\b = 8\N_a \O_\a,\quad \g^b_{\widetilde\a\widetilde\b} T_{ab}{}^{\widetilde\b} = 8 \N_a \Oh_{\widetilde\a}.
\label{nice}
\end{align}

The fermionic derivatives of $\O_\a$ and $\O_{\widetilde\a}$ are related to the RR field-strength superfield as
\begin{align}
\N_{\widetilde\a} \O_\a=\frac18(\g_a P \g^a)_{\a\widetilde\a},\quad \N_\a \Oh_{\widetilde\a}=-\frac18(\g_a P \g^a)_{\a\widetilde\a}. 
\label{nice2}
\end{align}
These equations come from the Bianchi identities involving $R_{\widetilde\a(\a\b)}{}^\b$ and $\Rh_{\a(\widetilde\a\widetilde\b)}{}^{\widetilde\b}$.

The torsion component $T_{abc}$ is related to $\Phi$. Mixing the Bianchi identity $R_{(\a\b\g)}{}^\r=0$ and the Bianchi identity involving $R_{\a\b ab}$, together with the Bianchi identity involving $\N_\a H_{abc}$ and the constraint $H_{abc}+T_{abc}=0$, we obtain the equation  
\begin{align}
R_{(\a\b} \d_{\g)}^\r  - \g_a^{\r\s} \g^a_{(\a\b} \N_{\g)}\O_\s  + \frac1{16} (\g^{ab})_{(\a}{}^\r \g^c_{\b\g)} T_{abc} = 0,
\label{Tfi}
\end{align}
where $R_{\a\b}=\N_{(\a}\O_{\b)}$. This implies 
\begin{align}\label{NNPhi}
    \N_\a\O_\b=\frac14\nabla_\alpha\nabla_\beta\Phi=\frac18\g^a_{\a\b}\N_a\Phi-\frac1{96}\g^{abc}_{\a\b}H_{abc}.
\end{align}
Similarly, the Bianchi identity $\Rh_{(\widetilde\a\widetilde\b\widetilde\g)}{}^{\widetilde\r}=0$, the Bianchi identities involving $\Rh_{\widetilde\a\widetilde\b ab}$ and $\Nh_{\widetilde\a}H_{abc}$, together with the constraint $H_{abc}-\Th_{abc}=0$ imply the equation
\begin{align}\label{NtildeNtildePhi}
   \N_{\widetilde\alpha}\Oh_{\widetilde\beta}=
   \frac14\nabla_{\widetilde\alpha}\nabla_{\widetilde\beta}\Phi=
   \frac18\g^a_{\widetilde\alpha\widetilde\beta}\N_a\Phi+\frac 1{96}\g^{abc}_{\widetilde\alpha\widetilde\beta}H_{abc}.
\end{align}

\subsection{Redefining the connections}
We now redefine the connections to have only one connection for vectors. This will be useful in the next section. As we reviewed above, the Berkovits-Howe constraints for type II supergravity imply that 
\begin{align}
    \O=E^\a\O_\a=\frac14E^\a\N_\a\Phi,\quad \Oh=E^{\widetilde\alpha}\Oh_{\widetilde\alpha}=\frac14E^{\widetilde\alpha}\N_{\widetilde\alpha}\Phi ,
\end{align}
and $\O_{ab}, \Oh_{ab}$ are related according to \cite{Chandia:2019paj}
\begin{align}
    \Oh_{cab} - \O_{cab} = - T_{cab}=H_{cab} ,\quad \Oh_{\a ab} - \O_{\a ab}=- T_{\a ab}  ,\quad \Oh_{\widetilde\a ab} - \O_{\widetilde\a ab}= \Th_{\widetilde\a ab},
\end{align}
or written in terms of forms 
\begin{align}\label{A3}
    \Oh_{ab}-\Omega_{ab}=F_{ab},
\end{align}
where 
\begin{align}
    F_{ab}=E^c H_{cab}-E^\alpha T_{\alpha ab} + 
    E^{\widetilde\alpha} \Th_{\widetilde\alpha ab}.
\end{align}
We now redefine the connections to have only one connection for vectors. The redefinition is
\begin{align}
    \O_\a{}^\b=\O'_\a{}^\b + x\d_\a^\b \O - \frac18(\g^{ab})_\a{}^\b F_{ab},\quad \Oh_{\widetilde\alpha}{}^{\widetilde\beta}=\Oh'_{\widetilde\alpha}{}^{\widetilde\beta}+x\d_{\widetilde\alpha}^{\widetilde\beta}\Oh+\frac18(\g^{ab})_{\widetilde\alpha}{}^{\widetilde\beta} F_{ab} ,
\end{align}
where $x$ is a free parameter that can be chosen later for convenience. From here, one obtains
\begin{align}
    &\O'_\a{}^\b=(1-x)\O\d_\a^\b+\frac14(\g^{ab})_\a{}^\b\left(\O_{ab}+\frac12 F_{ab}\right),\cr 
    &\Oh'_{\widetilde\alpha}{}^{\widetilde\beta}=(1-x)\Oh\d_{\widetilde\alpha}^{\widetilde\beta}+\frac14(\g^{ab})_{\widetilde\alpha}{}^{\widetilde\beta}\left(\Oh_{ab}-\frac12 F_{ab}\right).
\end{align}
Note that $\O_{ab}+\frac12 F_{ab}=\Oh_{ab}-\frac12 F_{ab}$ because (\ref{A3}). The new torsions can be calculated as follows 
\begin{align}
    &T'^\alpha=\nabla' E^\alpha = \nabla E^\alpha - x E^\alpha \Omega +\half E^{\beta}F_{\beta}{}^{\alpha},\\
    &T'^a =\nabla' E^a =\nabla E^a +\half E^b F_{b}{}^a,\\
    &T'^{\widetilde\alpha} =\nabla' E^{\widetilde\alpha}= \nabla E^{\widetilde\alpha}- x E^{\widetilde\alpha}\widehat\Omega -\half E^{\widetilde\beta}F_{\widetilde\beta}{}^{\widetilde\alpha}.
\end{align}
The components of $T'^{\alpha}$ are 
\begin{align}\label{Tnew1}
    &T'_{ab}{}^\a=T_{ab}{}^\a,\quad T'_{a\widetilde\beta}{}^\a = T_{a\widetilde\beta}{}^\a,\quad T'_{\widetilde{\gamma}\widetilde\beta}{}^\a=T_{\widetilde{\gamma}\widetilde\beta}{}^\a=0,\cr 
    &T'_{\g\b}{}^\a=\left(\frac52-x\right)\d^\a_{(\g}\O_{\b)}-2\g^a_{\g\b}(\g_a\O)^\a,\cr 
    &T'_{a\b}{}^\a=\frac18(\g^{bc})_\b{}^\a H_{abc},\quad T'_{\widetilde{\gamma}\b}{}^\a=\frac14(\g^{ab})_\b{}^\a(\g_{ab}\Oh)_{\widetilde{\gamma}} .
\end{align}
The components of $T'^{\widetilde\alpha}$ are 
\begin{align}\label{Tnew2}
    &T'_{ab}{}^{\widetilde\alpha}=T_{ab}{}^{\widetilde\alpha},\quad T'_{a\b}{}^{\widetilde\alpha} = T_{a\b}{}^{\widetilde\alpha},\quad T'_{\g\b}{}^{\widetilde\alpha}=T_{\g\b}{}^{\widetilde\alpha}=0,\cr 
   &T'_{\widetilde{\gamma}\widetilde\beta}{}^{\widetilde\alpha}=\left(\frac52-x\right)\d^{\widetilde\alpha}_{(\widetilde{\gamma}}\Oh_{\widetilde\beta)}-2\g^a_{\widetilde{\gamma}\widetilde\beta}(\g_a\Oh)^{\widetilde\alpha},\cr 
    &T'_{a\widetilde\beta}{}^{\widetilde\alpha}=-\frac18(\g^{bc})_{\widetilde\beta}{}^{\widetilde\alpha} H_{abc},\quad T'_{\g\widetilde\beta}{}^{\widetilde\alpha}=\frac14(\g^{ab})_{\widetilde\beta}{}^{\widetilde\alpha}(\g_{ab}\O)_\g .
\end{align}
And the components of $T'^a$ are 
\begin{align}\label{Tnew3}
    &T'_{cb}{}^a=T_{cb}{}^a+H_{cb}{}^a=0,\quad T'_{\a\b}{}^a=T_{\a\b}{}^a=-\g^a_{\a\b},\quad T'_{\widetilde{\alpha}\widetilde{\beta}}{}^a=T_{\widetilde{\alpha}\widetilde{\beta}}{}^a=-\g^a_{\widetilde{\alpha}\widetilde{\beta}} \cr 
    &T'_{\a b}{}^a=(\g_b{}^a\O)_\a,\quad T'_{\widetilde{\alpha} b}{}^a=(\g_b{}^a\Oh)_{\widetilde{\alpha}},\quad T'_{\a\widetilde{\beta}}{}^a=T_{\a\widetilde{\beta}}{}^a=0.
\end{align}
The new curvature two-form becomes
\begin{align}\label{Rnew}
    R'_{ab}=R_{ab}+\frac12\N F_{ab}+\frac14 F_a{}^c F_{cb},
\end{align}
when it is expressed in components, we obtain
\begin{align}\label{RprimeFromR}
    R'_{ABab}=R_{ABab}+\frac12\N_{[A}F_{B]ab}+\frac12 T_{AB}{}^C F_{Cab}-\frac14F_{[Aa}{}^c F_{B]cb} .
\end{align}

In particular,
\begin{align}\label{newcurvs}
    R'_{\a\b ab}=&\frac14(\g_{[a})_{\a\b}\N_{b]}\Phi+\frac38\g^c_{\a\b}\left(H_{abc}-6(\O\g_{abc}\O)\right) +\frac1{48}(\g_{abcde})_{\a\b}\left(H^{cde}-6(\O\g^{cde}\O)\right) ,\cr 
    R'_{\widetilde\alpha\widetilde\beta ab}=&\frac14(\g_{[a})_{\widetilde\alpha\widetilde\beta}\N_{b]}\Phi -\frac38\g^c_{\widetilde\alpha\widetilde\beta}\left(H_{abc}+6(\Oh\g_{abc}\Oh)\right) -\frac1{48}(\g_{abcde})_{\widetilde\alpha\widetilde\beta}\left(H^{cde}+6(\Oh\g^{cde}\Oh)\right) ,\cr
    R'_{\a\widetilde\beta ab}=&\frac18(\g_{ab}\g_c P\g^c)_{\a\widetilde\beta}+\frac18(\g^c P\g_c\g_{ab})_{\a\widetilde\beta}-(\g_{[a}P\g_{b]})_{\a\widetilde\beta}+(\g_{[a}{}^c\O)_\a (\g_{b]c}\Oh)_{\widetilde\beta} .
\end{align}

Interestingly, with this choice of connection, the RR fields and $H$ appear similarly in the torsions. We want to stress that the redefinition does not change how the BRST charge acts in vertex operators in the lowest order of $\alpha'$. We will drop all $'$s from torsions and curvatures in the following sections. The scaling connections $(\Omega_{\alpha},\widehat\Omega_{\widetilde\alpha})$, when they appear in expressions below,  are shorthand for $(\frac14\nabla_{\alpha }\Phi,\frac14\nabla_{\widetilde\alpha}\Phi)$ respectively.

\section{Superspace compactification}
\label{4Dsusy}

In this section, we will describe in more detail what we mean by superspace compactification described in the introduction. Using the the $\mathfrak{sl}(2,\mathbb{C})+\mathfrak{su}(4)$ decomposition the basic operators acting on general tensor superfields are $(\nabla_a,\nabla_i,\nabla_{\alpha I},\widetilde\nabla_{\alpha \widetilde I},\nabla_{\dot\alpha}{}^ I,\widetilde\nabla_{\dot\alpha}{}^{\widetilde I},M_{ab},M_{ai},M_{ij})$. The position of the index $\widetilde I$ depends on whether we are discussing Type IIA or IIB. For Type IIB $\widetilde\nabla_{\alpha \widetilde I}$ means $\widetilde\nabla_{\alpha I}$ and $\widetilde\nabla_{\dot\alpha}{}^{\widetilde I}$ means $\widetilde\nabla_{\dot\alpha}{}^{I}$ and for Type IIA $\widetilde\nabla_{\alpha \widetilde I}$ means 
$\widetilde\nabla_{\alpha}{}^I$ and $\widetilde\nabla_{\dot\alpha}{}^{\widetilde I}$ means 
$\widetilde\nabla_{\dot\alpha I}$. 

A central concept in compactifications is the existence of nowhere-vanishing spinors in the compactification manifold. This work will suppose two such spinors exist and promote them to bosonic, complex superfields. We will denote them $(\chi^I,\widetilde\chi^{\widetilde I})$. The constraints they satisfy will be derived throughout this paper. For Type IIB $\widetilde\chi^{\widetilde I}=\widetilde\chi^I$ and for Type IIA $\widetilde\chi^{\widetilde I}=\widetilde\chi_I$. Furthermore, they satisfy the reality conditions $\left(\chi^I\right)^\dagger=\chi_I$ and $\left(\widetilde\chi^{\widetilde I}\right)^\dagger=\widetilde\chi_{\widetilde I}$. We define the following covariant derivatives
\begin{align}
    \nabla_\alpha= \chi^I\nabla_{\alpha I} + \widetilde\chi^{\widetilde I}\widetilde\nabla_{\alpha \widetilde I},\quad \nabla_{\dot\alpha}=\chi_I\nabla_{\dot\alpha}{}^I+\widetilde\chi_{\widetilde I}\widetilde\nabla_{\dot\alpha}{}^{\widetilde I}.
\end{align}
The set of operators $(\nabla_\alpha,\nabla_{\dot\alpha},\nabla_{\alpha\dot\alpha}\equiv \{\nabla_\alpha,\nabla_{\dot\alpha}\},M_{ab})$ will be called \textit{four dimensional} or \textit{space-time operators}. The set composed by $(\nabla_i,M_{ij})$ will be called \textit{internal operators}. Finally, the set $(\Delta_\alpha,\Delta_{\dot\alpha},\Delta_{\alpha i},\Delta_{\dot\alpha i},\widetilde\Delta_{\alpha i},\widetilde\Delta_{\dot\alpha i},M_{ai})$ defined in (\ref{mixed_derivatives}) will be called \textit{mixed operators}. For the next definition, we will use $\circ$ in place of any combination of four-dimensional Lorentz indices $(a,\alpha,\dot\alpha)$ and $\bullet$ for any combination of local internal tangent space indices $(i, I)$. A \textit{four dimensional tensor superfield} ${\mathcal T}^\circ$ will satisfy
\begin{align}\label{fourDtensor}
    \nabla_i {\mathcal T}^\circ = M_{ij} {\mathcal T}^\circ=0.
\end{align}
Similarly, an \textit{internal tensor superfield} will satisfy 
\begin{align}
    \nabla_\alpha {\mathcal T}^\bullet =\nabla_{\dot\alpha}{\mathcal T}^\bullet= M_{ab}{\mathcal T}^\bullet=0.
\end{align}

Mixed operators will not be required to act trivially in any tensor superfield. In the same way, \textit{mixed tensor superfields} ${\mathcal T}^{\circ\bullet}$ will not be annihilated by any type of operator. The first example of an internal tensor superfield is the dilaton superfield $\Phi$. Since we want to preserve four-dimensional Poincar\'e or $SO(2,4)$ covariance, it is natural to assume that it is an internal superfield and satisfy 
\begin{align}\label{Nphi}
    \nabla_\alpha\Phi=\nabla_{\dot\alpha}\Phi=0.
\end{align}
We will also see how this condition appears again later. Note that using (\ref{OmPhi}) the equation (\ref{Nphi}) can be written as
\begin{align}\label{chiphi}
    \chi^I\O_{\a I}+\widetilde\chi^{\widetilde I} \Oh_{\a\widetilde I}=0.
\end{align}
Doing a Lorentz rotation with $M_{ai}$ we also have the following condition
\begin{align}\label{NphiDD}
    \chi^I\sigma_{IJ}^i\Omega_{\dot\alpha}{}^J +\widetilde\chi^{\widetilde I}\sigma^i_{\widetilde I\widetilde J}\widetilde\Omega_{\dot\alpha}{}^{\widetilde J}=0.
\end{align}
The commutator $[M_{ai}, M_{bj}]$ gives no further constraints if $\chi$ and $\widetilde\chi$ are proportional to each other, but it will give a new constraint otherwise. There are also the conjugates of the equations above. 

The spinor superfields $(\chi,\widetilde\chi)$ are further examples of internal superfields 
\begin{align}\label{internalChi}
    \nabla_\alpha\chi^I=\nabla_{\dot\alpha}\chi^I= \nabla_\alpha\widetilde\chi^{\widetilde I}=\nabla_{\dot\alpha}\widetilde\chi^{\widetilde I}=0.
\end{align}

Furthermore, we will require that the four-dimensional operators  satisfy an $AdS_4$ (or $d=4$ Poincar\'e) superalgebra
\begin{align}\label{susicalg1}
    &\{\nabla_\alpha,\nabla_\beta\}=  F M_{\a\b},\quad 
    \{\nabla_{\dot\alpha},\nabla_{\dot\beta}\}= - F^\dagger M_{\dot\alpha\dot\beta},\quad
    \{\nabla_\alpha,\nabla_{\dot\alpha}\}=\nabla_{\alpha\dot\alpha},\\ \label{susicalg2}
    &[\nabla_\alpha,\nabla_{\beta\dot\beta}]=-\frac12 \epsilon_{\alpha\beta}F \nabla_{\dot\beta},\quad
    [\nabla_{\dot\alpha},\nabla_{\beta\dot\beta}]=\frac12 \epsilon_{\dot\alpha\dot\beta}F^\dagger \nabla_{\beta},\\ \label{susicalg3}
    &[\nabla_{\alpha\dot\alpha},\nabla_{\beta\dot\beta}]=\frac{1}{2}|F|^2\left(\epsilon_{\dot\alpha\dot\beta}M_{\alpha\beta}+\epsilon_{\alpha\beta}M_{\dot\alpha\dot\beta}\right),
\end{align}
where $F$ is a bosonic complex superfield and $(M_{\alpha\beta},M_{\dot\alpha\dot\beta})$ are the four dimensional Lorentz generators in spinor basis. In the case of a flat four-dimensional compactification, we will require $F=0$. The third equation in (\ref{susicalg1}) should be interpreted as the definition of $\nabla_{\alpha\dot\alpha}$. Once the anti-commutators in (\ref{susicalg1}) are defined, the closure of the algebra fixes the remaining equations. For that to happen, it is necessary to impose that  $F$ is an internal superfield
\begin{align}
\label{susicalg4}
    \nabla_{\alpha} F=\nabla_{\dot\alpha} F=0.    
\end{align}

Usually $\nabla_{\alpha\dot\alpha}$ is defined with an $i$, which explains the unusual sign in (\ref{susicalg3}). One of our goals is to compute the superfield $F$ in terms of the torsions and curvatures implied by the Berkovits-Howe supergravity constraints presented in previous sections and $(\chi^I,\widetilde\chi^{\widetilde I})$. 

\subsection{Factorization and the Pure Spinor Equations}\label{factorization}

From the definitions above, it is natural to assume that space-time operators map space-time tensors to themselves and analogously to internal operators. There are possible nontrivial commutators among the super-translations in the two sets. Since $[\nabla_i,\nabla_\alpha]$ act trivially on both ${\mathcal T}^\circ$ and ${\mathcal T}^\bullet$ we will require that 
\begin{align}\label{factorizing}
    [\nabla_i,\nabla_\alpha]= 0,
\end{align}
The Bianchi identity with $[\nabla_i,\{\nabla_\alpha,\nabla_{\dot\alpha}\}]$ implies that $[\nabla_i,\nabla_{\alpha\dot\alpha}]$ also vanishes.  The condition (\ref{factorizing}) implies many expected restrictions, a few of which are already implied by (\ref{Nphi}). Expanding the left hand side of (\ref{factorizing}) we have
\begin{align}\label{NiNa}
    [\nabla_i,\nabla_\alpha]=&\;(\nabla_i\chi^I)\nabla_{\alpha I}+(\nabla_i\widetilde\chi^{\widetilde I})\widetilde\nabla_{\alpha\widetilde I}+\chi^I[\nabla_i,\nabla_{\alpha I}]+\widetilde\chi^{\widetilde I}[\nabla_i,\widetilde\nabla_{\alpha \widetilde I}]\\
    =&\; \left( \left(\nabla_i\chi^I -\frac18\chi^J(\s_{jk})_J{}^I H_{ijk}\right)\delta^\beta_\alpha -\widetilde\chi^{\widetilde J}(\s_i)_{\widetilde J\widetilde K} P^{\b I}{}_\a{}^{\widetilde K}-\frac18 (\sigma^{ab})_\alpha{}^\beta\chi^I H_{iab}\right) \nabla_{\beta I}\cr 
    +&\; \left(\left(\nabla_i\widetilde\chi^{\widetilde I}+\frac18\widetilde\chi^{\widetilde J}(\s_{jk})_{\widetilde J}{}^{\widetilde I} H_{ijk}\right)\delta^\beta_\alpha +\chi^J(\s_i)_{JK} P_\a{}^{K\b\widetilde I}+\frac18 (\sigma^{ab})_\alpha{}^\beta\widetilde\chi^{\widetilde I} H_{iab} \right
    )\widetilde\nabla_{\beta\widetilde I}\cr 
    -&\left(\chi^I(\s_i)_{IJ}\O^{\dot\b J}+\widetilde\chi^{\widetilde I}(\s_i)_{\widetilde I\widetilde J}\Oh^{\dot\b\widetilde J}\right)(\s^a)_{\a\dot\b}\N_a 
    -\left(\chi^I(\s_{ij})_I{}^J\O_{\a J}+\widetilde\chi^{\widetilde I}(\s_{ij})_{\widetilde I}{}^{\widetilde J}\Oh_{\a\widetilde J}\right)\N_j\cr
    -&\;\left( \frac18\chi^I (\sigma_j)_{IJ}(\sigma^a)_\alpha{}^{\dot\beta}H_{ija} -\widetilde\chi^{\widetilde I} (\sigma_i)_{\widetilde I\widetilde K}P^{\dot\beta}{}_{J}{}_\alpha{}^{\widetilde K} \right)\nabla_{\dot\beta}{}^J\cr 
    -&\;\left( -\frac18\widetilde\chi^{\widetilde I}(\sigma_j)_{\widetilde I\widetilde J}(\sigma^a)_\alpha{}^{\dot\beta}H_{ija}+\chi^I(\sigma_i)_{IJ}P^{\alpha J\dot\beta}{}_{\widetilde J}\right)\widetilde\nabla_{\dot\beta}{}^{\widetilde J}\cr \label{factorizing2}
    +&\;\frac12\left(\chi^I R_{i\alpha I}{}^{\underline{ab}}+\widetilde\chi^{\widetilde I} R_{i\alpha\widetilde I}{}^{\underline{ab}}\right)M_{\underline{ab}}.
\end{align}
We will now discuss the implication of each line of the second equality in (\ref{NiNa}).
The last line provides constraint equations involving the curvature, namely
\begin{align}\label{curvXR}
    \chi^I R_{i\alpha I}{}^{\underline{ab}}+\widetilde\chi^{\widetilde I} R_{i\alpha\widetilde I}{}^{\underline{ab}}=0.
\end{align}
This condition has mass dimension $3/2$; its first non-vanishing contribution in the covariant $\theta$-expansion has dimension two. 
The first term in the third line, with $\N_a$, 
vanish  (\ref{NphiDD}). The second term, with $\N_j$, implies that 
\begin{align}
    \chi^I(\s_{ij})_I{}^J\O_{\a J}+\widetilde\chi^{\widetilde I}(\s_{ij})_{\widetilde I}{}^{\widetilde J}\Oh_{\a\widetilde J}=0
\end{align}
This is not an independent condition if $\widetilde\chi$ is proportional to $\chi$, as explained after (\ref{NphiDD}). The fourth and fifth lines imply relations between $H$ and the RR field strengths. However, demanding that any background is an internal superfield, we should also require that 
\begin{align}
    H_{ijc}=P^{\alpha I\dot\beta}{}_{\widetilde J}=0.
\end{align}
This requirement also demands that 
\begin{align}
    H_{iab}=0, 
\end{align}
and that 
\begin{align}
    P^{\alpha I}{}_\beta{}^{\widetilde J}= \delta^\alpha{}_{\beta}F^{I\widetilde J}, 
\end{align}
since this is sufficient for $[M_{ab},P^{\alpha I}{}_\beta{}^{\widetilde J}]=0$.

Finally, the factorization condition (\ref{factorization}) implies that the normalizable spinors satisfy 
\begin{align}\label{pureSpinorEqs}
    &\N_i\chi^I-\frac18\chi^J(\s_{jk})_J{}^I H_{ijk}+\widetilde\chi^{\widetilde J} (\s_i)_{\widetilde J\widetilde K} F^{I\widetilde K}=0,\\ 
    &\N_i\widetilde\chi^{\widetilde I}+\frac18\widetilde\chi^{\widetilde J}(\s_{jk})_{\widetilde J}{}^{\widetilde I} H_{ijk}+\chi^J (\s_i)_{JK} F^{K\widetilde I}=0.
\end{align}

Defining bi-spinors superfields with $(\chi,\widetilde\chi)$, \textit{e.g.} $\Psi^{IJ}=\chi^I\chi^J$, they will satisfy the tangent space version of the so-called \textit{pure spinor equations} \cite{Grana:2004bg,Grana:2005sn} up to a scale factor in $(\chi,\widetilde\chi)$. The exponential of the dilaton that appears in the literature is contained in $F^{I\widetilde J}$, since Ramond-Ramond vertex operators have this factor \cite{Polyakov:1995bn}. However, a fundamental difference exists; (\ref{pureSpinorEqs}) is a superspace condition. It implies higher mass dimension equations in the covariant $\theta$ expansion.

\subsection{The cosmological constant}

We now compute the algebra (\ref{susicalg1}) and imposing its form will imply further constraints on the background geometry. Recall the definition
\begin{align}
    \N_\a=\chi^I\N_{\a I}+\widetilde\chi^{\widetilde I}\N_{\a\widetilde I}.
\end{align}
Consider $\{\N_\a,\N_\b\}$. It becomes
\begin{align}
    \{\N_\a,\N_\b\}&=\chi^I\chi^J\{\N_{\a I},\N_{\b J}\}+\widetilde\chi^{\widetilde I}\widetilde\chi^{\widetilde J}\{\N_{\a\widetilde I},\N_{\b\widetilde J}\}\cr
    &+\chi^I\widetilde\chi^{\widetilde J}\left( \{\N_{\a I},\N_{\b\widetilde J}\}+\{\N_{\b I},\N_{\a\widetilde J}\}\right) .
\end{align}
We use the torsion $T'$ and curvature $R'$ components defined in the previous section (equations (\ref{Tnew1}), (\ref{Tnew2}), (\ref{Tnew3}), (\ref{Rnew})) to compute the anticommutators. We obtain
\begin{align}\label{antcom1}
    &\{\N_\a,\N_\b\}=\left(x-\frac52\right)\chi^I\chi^J \O_{(\a I}\N_{\b)J}+\left(x-\frac52\right)\widetilde\chi^{\widetilde I}\widetilde\chi^{\widetilde J}\Oh_{(\a\widetilde I}\N_{\b)\widetilde J} \cr 
    &+\chi^I\widetilde\chi^{\widetilde J}\left( \O_{(\a I}\N_{\b)\widetilde J}+\Oh_{(\a\widetilde J}\N_{\b) I}-\frac14(\s_{ij})_I{}^K(\s_{ij})_{\widetilde J}{}^{\widetilde L}\left( \O_{(\a K}\N_{\b)\widetilde L}+\Oh_{(\a\widetilde L}\N_{\b)K} \right) \right) \cr
    &+\frac12 U_{\alpha\beta}{}^{\underline{ab}} M_{\underline{ab}},
\end{align}
where 
\begin{align}
    U_{\alpha\beta}{}^{\underline{ab}}=\chi^I\chi^J R_{\a I\b J}{}^{\underline{ab}} +\widetilde\chi^{\widetilde I}\widetilde\chi^{\widetilde J} R_{\a\widetilde I\b\widetilde J}{}^{\underline{ab}} +\chi^I\widetilde\chi^{\widetilde J} R_{(\a I\b)\widetilde J}{}^{\underline{ab}}.
\end{align}
The vanishing of the derivative terms imposes that 
\begin{align}\label{noderiv1}
    &\left(x-\frac72\right)\chi^J\chi^I\Omega_{\alpha I}-\frac14(\chi\sigma_{ij})^J(\widetilde\chi\sigma_{ij}\widehat\Omega_{\alpha})=0,\\
    \label{noderiv2}
    &\left(x-\frac72\right)\widetilde\chi^{\widetilde J}\widetilde\chi^{\widetilde I}\widehat\Omega_{\alpha \widetilde I}-\frac14(\widetilde\chi\sigma_{ij})^{\widetilde J}(\chi\sigma_{ij}\Omega_{\alpha})=0,
\end{align}
together with their complex conjugates. We have used (\ref{chiphi}) to simplify the expressions above. We will discuss solving this in section \ref{possibilities}. 
The term with $M_{ab}$ will give the $F$ of the first anti-commutator in (\ref{susicalg1}). The terms with $M_{a i}$ and $M_{ij}$ should vanish
\begin{align}\label{Uai}
   &U_{\alpha\beta}{}^{ai}=\chi^I\chi^J R_{\a I\b J}{}^{{ai}} +\widetilde\chi^{\widetilde I}\widetilde\chi^{\widetilde J} R_{\a\widetilde I\b\widetilde J}{}^{{ai}} 
+\chi^I\chi^{\widetilde J} R_{\a I\b\widetilde J}{}^{{ai}}+\chi^I\widetilde\chi^{\widetilde J}R_{\b I\a\widetilde J}{}^{{ai}}=0 ,\\ \label{Uij}
    &U_{\alpha\beta}{}^{ij}=\chi^I\chi^J R_{\a I\b J}{}^{{ij}} +\widetilde\chi^{\widetilde I}\widetilde\chi^{\widetilde J} R_{\a\widetilde I\b\widetilde J}{}^{{ij}} 
+\chi^I\chi^{\widetilde J} R_{\a I\b\widetilde J}{}^{{ij}}+\chi^I\widetilde\chi^{\widetilde J}R_{\b I\a\widetilde J}{}^{{ij}}=0 .
\end{align}
With the assumptions above, we have 
 \begin{align}\label{result1}
    \{\N_\a,\N_\b\}&=\frac12\left(\chi^I\chi^J R_{\a I\b J}{}^{{ab}}  + \widetilde\chi^{\widetilde I}\chi^{\widetilde J} R_{\a\widetilde I\b\widetilde J}{}^{{ab}} 
    +\chi^I\widetilde\chi^{\widetilde J} ( R_{\a I\b\widetilde J}{}^{{ab}}+R_{\b I\a\widetilde J}{}^{{ab}})\right) M_{{ab}}. 
\end{align}
And finally, we obtain that 
\begin{align}\label{cosmologicalConstant}
    F=\frac1{24}(\sigma_{ab})^{\alpha\beta} \left(\chi^I\chi^J R_{\a I\b J}{}^{{ab}}  + \widetilde\chi^{\widetilde I}\chi^{\widetilde J} R_{\a\widetilde I\b\widetilde J}{}^{{ab}} 
    +\chi^I\widetilde\chi^{\widetilde J} ( R_{\a I\b\widetilde J}{}^{{ab}}+R_{\b I\a\widetilde J}{}^{{ab}})\right).
\end{align}
The coefficient of the four-dimensional Lorentz generator $M_{ab}$ is symmetric in $\alpha\beta$, so there are no further constraints coming from (\ref{result1}). Later, it will be checked if this is the same internal superfield in the first anti-commutator (\ref{susicalg2}).

\subsection{Closure of the algebra}
The next step is to define $\N_{\a\dot\a}=\{\N_\a,\N_{\dot\a}\}$. Using $\N_{\dot\a}\chi^I=\N_{\dot\a}\widetilde\chi^{\widetilde I}=\N_\a\chi_I=\N_\a\widetilde\chi_{\widetilde I}=0$ we obtain 
\begin{align}   \label{fourDderivative}
    \N_{\a\dot\a}&=|\chi|^2(\s^a)_{\a\dot\a}\N_a+|\widetilde\chi|^2(\s^a)_{\a\dot\a}\N_a+\frac12 U_{\a\dot\a}{}^{\underline{ab}}M_{\underline{ab}}\cr
    &+S_{\alpha J}\nabla_{\dot\alpha}{}^J +\widetilde S_{\alpha\widetilde J}\widetilde\nabla_{\dot\alpha}{}^{\widetilde J}+S_{\dot\alpha}{}^{J}\nabla_{\alpha J} +\widetilde S_{\dot\alpha}{}^{\widetilde J}\widetilde\nabla_{\alpha\widetilde J}
\end{align}
where 
\begin{align}
    &S_{\alpha J}=(x-5/2)\chi^I\chi_J\O_{\a I}+4|\chi|^2\Omega_{\alpha J}+(\chi\sigma_i)_J(\widetilde\chi\sigma_i\widehat\Omega_{\alpha})-\frac14 (\chi\sigma_{ij})_J(\widetilde\chi\sigma_{ij}\widehat\Omega_{\alpha}),\\
    &\widetilde S_{\alpha\widetilde J}=(x-5/2)\widetilde\chi^I\widetilde\chi_J\widehat\O_{\a \widetilde I}+4|\widetilde\chi|^2\widehat\Omega_{\alpha \widetilde J}+(\widetilde\chi\sigma_i)_{\widetilde J}(\chi\sigma_i\Omega_{\alpha})-\frac14 (\widetilde\chi\sigma_{ij})_{\widetilde J}(\chi\sigma_{ij}\Omega_{\alpha}),\\
    &S_{\dot\a}{}^J=(x-5/2)\chi^J\chi_I\O_{\dot\a}{}^I+4|\chi|^2\O_{\dot\a}{}^J+(\chi\s_i)^J(\widetilde\chi\s_i\Oh_{\dot\a})-\frac14(\chi\s_{ij})^J(\widetilde\chi\s_{ij}\Oh_{\dot\a}),\\
    &\widetilde S_{\dot\a}{}^{\widetilde J}=(x-5/2)\widetilde\chi^{\widetilde J}\widetilde\chi_{\widetilde I}\Oh_{\dot\a}{}^{\widetilde I}+4|\widetilde\chi|^2\Oh_{\dot\a}{}^{\widetilde J}+(\widetilde\chi\s_i)^{\widetilde J}(\chi\s_i\O_{\dot\a})-\frac14(\widetilde\chi\s_{ij})^{\widetilde J}(\chi\s_{ij}\O_{\dot\a}),
    \end{align}
and,
\begin{align}\label{Uis}
    &U_{\a\dot\a}{}^{\underline{ab}}=\chi^I\chi_J R_{\a I\dot\a^J}{}^{\underline{ab}}+\widetilde\chi^{\widetilde I}\widetilde\chi_{\widetilde J} R_{\a\widetilde I\dot\a^{\widetilde J}}{}^{\underline{ab}}+\chi^I\widetilde\chi_{\widetilde J} R_{\a I\dot\a^{\widetilde J}}{}^{\underline{ab}}+\widetilde\chi^{\widetilde I}\chi_J R_{\a\widetilde I\dot\a^J}{}^{\underline{ab}} .
\end{align}
We will not impose that any of these terms vanish; the whole expression (\ref{fourDderivative}) defines $\nabla_{\alpha\dot\alpha}$.

Now we compute $[\N_\a,\N_{\b\dot\b}]$ which should be equal to $-\frac12 \e_{\a\b} F\N_{\dot\b}$, where $F$ should be equal to the result given in (\ref{cosmologicalConstant}). The curvatures in (\ref{Uis}) are given in (\ref{newcurvs}). The computation can be broken down into a few smaller steps. First, we can compute 
\begin{align}
   \frac12[\N_\a,U_{\b\dot\b}{}^{\underline{ab}}M_{\underline{ab}}]&=(\N_\a U_{\b\dot\b}{}^{\underline{ab}})M_{\underline{ab}}+\frac14 U_{\b\dot\b}{}^{ab}(\s_{ab})_\a{}^\g \N_{\g}\cr &
   +\frac12 U_{\b\dot\b}{}^{ai} (\s_a)_{\a\dot\g}(\s^i)_{IJ}\chi^I\N^{\dot\g J}+\frac12 U_{\b\dot\b}{}^{ai} (\s_a)_{\a\dot\g}(\s^i)_{\tilde I\tilde J}\tilde\chi^{\tilde I}\tilde\N^{\dot\g \tilde J}.
\end{align}
Note that $\nabla_\alpha$ and $\nabla_{\dot\alpha}$ should annihilate any background field. Next, we add the contribution coming only from $\nabla_a$
\begin{align}
   [\N_\a,\N_a]&=-(\N_a\chi^I)\N_{\a I}-(\N_a\tilde\chi^{\tilde I})\N_{\a\tilde I}\cr 
   &+\chi^I\left( (\s_a{}^b)_\a{}^\g\O_{\g I}\N_b+(\s_a)_{\a\dot\g}(\s_i)_{IJ}\O^{\dot\g J}\N_i+(\s_a)_{\a\dot\g} P^{\dot\g}{}_I{}^{\dot\r}{}_{\tilde J} \N_{\dot\r}{}^{\tilde J}\right.\cr
   &\left.+(x-1)\N_a\O_{\a I}{\mathcal S}+\frac12 R_{\a Ia}{}^{\underline{bc}} M_{\underline{bc}} \right)\cr
   &+\tilde\chi^{\tilde I}\left( (\s_a{}^b)_\a{}^\g\Oh_{\g\tilde I}\N_b+(\s_a)_{\a\dot\g}(\s_i)_{\tilde I\tilde J}\widehat\O^{\dot\g \tilde J}\N_i-(\s_a)_{\a\dot\g} P^{\dot\r}{}_J{}^{\dot\g}{}_{\tilde I} \N_{\dot\r}{}^{J}\right.\cr 
   &\left.+(x-1)\N_a\Oh_{\a\tilde I}\tilde{\mathcal S}+\frac12 R_{\a\tilde Ia}{}^{\underline{bc}} M_{\underline{bc}} \right), 
\end{align}
where $({\mathcal S},\tilde{\mathcal S})$ are the left and right scale generators. Note that 
\begin{align}
   &\N_a\O_{\a I}=\frac14\{\N_a,\N_{\a I}\}\Phi=-\frac14(\s_a)_{\a\dot\g}(\s_i)_{IJ}\O^{\dot\g J} \N_i\Phi-\frac14(\s_a)_{\a\dot\r} P^{\dot\r}{}_I{}^{\dot\g}{}_{\tilde J} \N_{\dot\g}{}^{\tilde J}\Phi,\\
   &\N_a\Oh_{\b\tilde I}=\frac14\{\N_a,\N_{\b\tilde I}\}\Phi 
   =-\frac14(\s_a)_{\b\dot\g}(\s_i)_{\tilde I\tilde J}\Oh^{\dot\g\tilde J} \N_i\Phi+\frac14(\s_a)_{\b\dot\r} P^{\dot\g}{}_J{}^{\dot\r}{}_{\tilde I} \N_{\dot\g}{}^{J}\Phi .
\end{align}
The final step is to compute $[\nabla_\alpha,S_{\beta J}\nabla_{\dot\beta}{}^J +\widetilde S_{\beta\widetilde J}\widetilde\nabla_{\dot\beta}{}^{\widetilde J}+S_{\dot\beta}{}^{J}\nabla_{\beta J} +\widetilde S_{\dot\beta}{}^{\widetilde J}\widetilde\nabla_{\beta\widetilde J}]$ which is done in Appendix \ref{NablaSComputation}. The final result is organized as 
\begin{align}\label{FinalNablaAlphaNablaA}
    [\nabla_\alpha,\nabla_{\beta\dot\beta}]&=\Psi_{\alpha\beta\dot\beta}{}^{\dot\g}{}_I\nabla_{\dot\gamma}{}^I+\widetilde\Psi_{\alpha\beta\dot\beta}{}^{\dot\gamma}{}_{\widetilde I}\widetilde\nabla_{\dot\gamma}{}^{\widetilde I}
    +W^a_{\alpha\beta\dot\beta}\nabla_a +X^i_{\alpha\beta\dot\beta}\nabla_i\cr  
    &+ Y_{\alpha\beta\dot\beta}{}^{\gamma I}\nabla_{\gamma I} +\widetilde Y_{\alpha\beta\dot\beta}{}^{\gamma \widetilde I}\widetilde\nabla_{\gamma\widetilde I} +Z_{\alpha\beta\dot\beta}{}^{\underline{ab}} M_{\underline{ab}}+
    Z^{\mathcal S}_{\alpha\beta\dot\beta}{\mathcal S}+
     Z^{\widetilde{\mathcal S}}_{\alpha\beta\dot\beta}\widetilde{\mathcal S},
\end{align}
where\footnote{The expressions below do not include terms with $\Omega^2$, $\widetilde\Omega^2$ and $\Omega\widetilde\Omega$. These terms are not included since they need at least two Grassmann odd derivatives for a non-vanishing contribution in the covariant $\theta$ expansion.}
\begin{align}\label{Psitensor}
    \Psi_{\alpha\beta\dot\beta}{}^{\dot\g}{}_I&=-\frac12 U_{\b\dot\b}{}^{ai} (\s_a)_{\a}{}^{\dot\g}\chi^J(\s^i)_{JI} -2\e_{\a\b}(|\chi|^2+|\widetilde\chi|^2)\widetilde\chi^{\widetilde J} \delta^{\dot\gamma}_{\dot\beta}F_{I\widetilde J}    \cr
    &+\e_{\a\b}\left(|\chi|^2(\chi\s_i)_I\N_i\Phi-\frac14(\chi\s_i)_I(\widetilde\chi\s_j\s_i\widetilde\chi)\N_j\Phi +\frac1{16}(\chi\s_{ij})_I(\widetilde\chi\s_{ijk}\widetilde\chi) \N_k\Phi\right)\d_{\dot\b}^{\dot\g} ,
\end{align}
\begin{align}\label{Psitildetensor}
    \widetilde\Psi_{\alpha\beta\dot\beta}{}^{\dot\gamma}{}_{\widetilde I}&=-\frac12 U_{\b\dot\b}{}^{ai} (\s_a)_{\a}{}^{\dot\g}\widetilde\chi^{\widetilde J}(\s^i)_{\widetilde J\widetilde I}-2\e_{\a\b}(|\chi|^2+|\widetilde\chi|^2)\chi^J\delta^{\dot\gamma}_{\dot\beta} F_{J\widetilde I} \cr 
    &+\e_{\a\b}\left(|\widetilde\chi|^2(\widetilde\chi\s_i)_{\widetilde I}\N_i\Phi-\frac14(\widetilde\chi\s_i)_{\widetilde I}(\chi\s_j\s_i\chi)\N_j\Phi +\frac1{16}(\widetilde\chi\s_{ij})_{\widetilde I}(\chi\s_{ijk}\chi) \N_k\Phi\right)\d_{\dot\b}^{\dot\g} ,
\end{align}
\begin{align}\label{Wtensor}
    W^a_{\alpha\beta\dot\beta}&=-(S_{\b I}\chi^I+\widetilde S_{\b\widetilde I}\widetilde\chi^{\widetilde I})(\s^a)_{\a\dot\b},
\end{align}
\begin{align}\label{Xtensor}
    X^i_{\alpha\beta\dot\beta}&=-\e_{\a\b}(S_{\dot\b}{}^J\chi^I(\s^i)_{IJ}+\widetilde S_{\dot\b}{}^{\widetilde J}\widetilde\chi^{\widetilde I}(\s^i)_{\widetilde I\widetilde J}),
\end{align}
\begin{align}\label{Ytensor}
    Y_{\alpha\beta\dot\beta}{}^{\gamma I}&=\frac14 U_{\b\dot\b}{}^{ab}(\s_{ab})_\a{}^\g \chi^I+\delta^\gamma_\alpha\left( \frac12 U_{\beta\dot\beta}{}^{ij}(\chi\sigma_{ij})^I+(S\cdot\nabla)_{\beta\dot\beta}\chi^I\right) ,
\end{align}
\begin{align}\label{Ytildetensor}
    \widetilde Y_{\alpha\beta\dot\beta}{}^{\gamma \widetilde I}&=\frac14 U_{\b\dot\b}{}^{ab}(\s_{ab})_\a{}^\g \widetilde\chi^{\widetilde I}+\delta^\gamma_\alpha\left( \frac12 U_{\beta\dot\beta}{}^{ij}(\widetilde\chi\sigma_{ij})^{\widetilde I}+(S\cdot\nabla)_{\beta\dot\beta}\widetilde\chi^{\widetilde I}\right),
\end{align}
\begin{align}\label{Zlorentztensor}
    Z_{\alpha\beta\dot\beta}{}^{\underline{ab}}&=\frac12(|\chi|^2+|\widetilde\chi|^2)(\s^a)_{\b\dot\b}\left(\chi^I R_{\a Ia}{}^{\underline{ab}}+\widetilde\chi^{\widetilde I}R_{\a\widetilde Ia}{}^{\underline{ab}}\right)+\frac12 \nabla_\alpha U_{\beta\dot\beta}{}^{\underline{ab}},
\end{align}
\begin{align}\label{ZStensor}
    Z^{\mathcal S}_{\alpha\beta\dot\beta}&=\left(\frac{1-x}{2}\right)\e_{\a\b}(|\chi|^2+|\widetilde\chi|^2)\chi^I\left((\s_i)_{IJ}\O_{\dot\b}{}^J\N_i\Phi+P_{\dot\b I}{}^{\dot\g}{}_{\widetilde J}\N_{\dot\g}{}^{\widetilde J}\Phi\right)\cr 
    &+\left(\frac{1-x}{4}\right)\e_{\a\b} S_{\dot\b}{}^J\chi^I(\s_i)_{IJ}\N_i\Phi     \cr
    &+(1-x)\left(S_{\b J}\widetilde\chi^{\widetilde I}\N_{\a\widetilde I}\O_{\dot\b}{}^J+\widetilde S_{\b\widetilde J}\chi^I\N_{\dot\b}{}^{\widetilde J}\O_{\a I}+S_{\dot\b}{}^J\widetilde\chi^{\widetilde I}\N_{\a\widetilde I}\O_{\b J}+\widetilde S_{\dot\b}{}^{\widetilde J}\chi^I\N_{\b\widetilde J}\O_{\a I}\right),
\end{align}
\begin{align}\label{ZStildetensor}
    Z^{\widetilde{\mathcal S}}_{\alpha\beta\dot\beta}&=\left(\frac{1-x}{2}\right)\e_{\a\b}(|\chi|^2+|\widetilde\chi|^2)\widetilde\chi^{\widetilde I}\left((\s_i)_{\widetilde I\widetilde J}\Oh_{\dot\b}{}^{\widetilde J}\N_i\Phi-P^{\dot\g}{}_{J\dot\b\widetilde I}\N_{\dot\g}{}^{J}\Phi\right)\cr  
    &+\left(\frac{1-x}{4}\right)\e_{\a\b}\widetilde S_{\dot\b}{}^{\widetilde J}\widetilde\chi^{\widetilde I}(\s_i)_{\widetilde I\widetilde J}\N_i\Phi     \cr
    &+(1-x)\left(S_{\b J}\widetilde\chi^{\widetilde I}\N_{\dot\b}{}^J\Oh_{\a\widetilde I}+\widetilde S_{\b\widetilde J}\chi^I\N_{\a I}\Oh_{\dot\b}{}^{\widetilde J}+S_{\dot\b}{}^J\widetilde\chi^{\widetilde I}\N_{\b J}\Oh_{\a\widetilde I}+\widetilde S_{\dot\b}{}^{\widetilde J}\chi^I\N_{\a I}\Oh_{\b\widetilde J}\right) .
\end{align}
We used the notation 
\begin{align}
    (S\cdot\nabla)_{\alpha\dot\alpha}=S_{\alpha J}\nabla_{\dot\alpha}{}^J +\widetilde S_{\alpha\widetilde J}\widetilde\nabla_{\dot\alpha}{}^{\widetilde J}+S_{\dot\alpha}{}^{J}\nabla_{\alpha J} +\widetilde S_{\dot\alpha}{}^{\widetilde J}\widetilde\nabla_{\alpha\widetilde J},
\end{align}
to have a shorter expression for the $Y$ tensors. Except for the first two tensors above, all others should vanish. The last three tensors have a mass dimension $3/2$, and the first nontrivial condition they imply has a mass dimension of $2$. Of course, there is the obvious choice of fixing $x=1$. But this is not the most convenient for a simple example. For the $Y$ tensors, their $\theta^0$ component implies no four-dimension Lorentz breaking background. The vanishing of $W$ and $X$ may imply interesting conditions that will be discussed later. 

The necessary conditions for the $\Psi$ tensors imply that 
\begin{align}\label{defineV}
    U_{\beta\dot\beta}{}^{ai}=V^i\sigma^a_{\beta\dot\beta},
\end{align}
for some internal superfield $V^i$. Substituting this in $\Psi$ and imposing (\ref{susicalg2}) we have the following equations
\begin{align}\label{secondF}
    -\frac12\chi_I F&= V^i\sigma^i_{JI}\chi^J -2(|\chi|^2+|\widetilde\chi|^2)\widetilde\chi^{\widetilde J}F_{I\widetilde J} +|\chi|^2(\chi\s_i)_I\N_i\Phi \cr 
    &-\frac14(\chi\s_i)_I(\widetilde\chi\s_j\s_i\widetilde\chi)\N_j\Phi +\frac1{16}(\chi\s_{ij})_I(\widetilde\chi\s_{ijk}\widetilde\chi) \N_k\Phi  
\end{align}
\begin{align}
    -\frac12\widetilde\chi_{\widetilde I} F&= V^i\sigma^i_{\widetilde J\widetilde I}\widetilde\chi^{\widetilde J} -2(|\chi|^2+|\widetilde\chi|^2)\chi^{ J}F_{J\widetilde I} +|\widetilde\chi|^2(\widetilde\chi\s_i)_{\widetilde I}\N_i\Phi \cr 
    &-\frac14(\widetilde\chi\s_i)_{\widetilde I}(\chi\s_j\s_i\chi)\N_j\Phi +\frac1{16}(\widetilde\chi\s_{ij})_{\widetilde I}(\chi\s_{ijk}\chi) \N_k\Phi 
\end{align}
Finally we calculate $F$ from the equations above contracting with $\chi$ and $\widetilde\chi$
respectively 
\begin{align}\label{closure1}
    F&= 4\left(1+\frac{|\widetilde\chi|^2}{
    |\chi|^2}\right)\chi^IF_{I\widetilde J}\widetilde\chi^{\widetilde J}-\frac{1}{8|\chi|^2}(\chi\s_{ij}\chi)(\widetilde\chi\s_{ijk}\widetilde\chi) \N_k\Phi\\ \label{closure2}
    &=4\left(1+\frac{|\chi|^2}{
    |\widetilde\chi|^2}\right)\chi^IF_{I\widetilde J}\widetilde\chi^{\widetilde J}-\frac{1}{8|\widetilde\chi|^2}(\widetilde\chi\s_{ij}\widetilde\chi)(\chi\s_{ijk}\chi) \N_k\Phi.
\end{align}
An $AdS_4$ solution is only possible if the above equations are equal to (\ref{cosmologicalConstant}).

\subsection{Summary}
Since this section describes numerous scattered conditions, discussing the main assumptions and consequences is convenient. The starting assumption is that it should be possible to construct a four-dimensional superspace covariant derivative algebra (\ref{susicalg1},\ref{susicalg2},\ref{susicalg3}) from the ten-dimensional one. We also impose that the four-dimensional operators should annihilate internal superfields. In particular, the dilaton superfield should satisfy 
\begin{align*}
\nabla_\alpha\Phi=0.    
\end{align*}
The four-dimensional algebra imposes various conditions, and we pay particular attention to the internal superfield $F$, whose first component corresponds to the cosmological constant. We found different expressions for it: (\ref{cosmologicalConstant}), (\ref{closure1}), and (\ref{closure2}). If they are all equal, the background will be consistent with four-dimensional supersymmetry. This will imply restrictions on the various possible bosonic background fields. Finally, the factorization discussed in the introduction also implies a familiar equation for the normalizable spinors 
\begin{align}
     &\N_i\chi^I-\frac18\chi^J(\s_{jk})_J{}^I H_{ijk}+\widetilde\chi^{\widetilde J} (\s_i)_{\widetilde J\widetilde K} F^{I\widetilde K}=0,\\ 
    &\N_i\widetilde\chi^{\widetilde I}+\frac18\widetilde\chi^{\widetilde J}(\s_{jk})_{\widetilde J}{}^{\widetilde I} H_{ijk}+\chi^J (\s_i)_{JK} F^{K\widetilde I}=0.
\end{align}
In the next section, we will discuss the implications of the abovementioned constraints. 

\section{A simple flux background}
\label{possibilities}

Let us now discuss a simple solution for the superspace constraints we derived \cite{Grana:2005sn, Grana:2005jc}. We are going to assume Type IIB supergravity and that the whole normalizable spinor superfields are equal
\begin{align}\label{chiequalchitilde}
    \chi^I=\widetilde\chi^I. 
\end{align}
With this assumption, we can see that (\ref{closure1}) is equal to (\ref{closure2}) as required. Furthermore, using (\ref{chiphi}) and (\ref{NphiDD}) and some identities in Appendix \ref{identities}, one can see that the $W$ and $X$ tensors in (\ref{Wtensor}) and (\ref{Xtensor}) are zero. The same can be shown for (\ref{noderiv1}) and (\ref{noderiv2}) choosing $x=5$. Another simplification is that with the choice $\chi^I=\widetilde\chi^I$ all contributions from $H_{ijk}$ in the tensors $U_{\alpha\beta}{}^{\underline{ab}}$, $U_{\dot\alpha\dot\beta}{}^{\underline{ab}}$ and $U_{\alpha\dot\beta}{}^{\underline{ab}}$ drop. 

With the assumption (\ref{chiequalchitilde}) the pure spinor equations (\ref{pureSpinorEqs}) imply that 
\begin{align}
    \frac14 \chi^J(\sigma_{jk})_J{}^I H_{ijk}+\chi^J(\sigma_i)_{JK}\left(F^{KI}-F^{IK}\right)=0.
\end{align}
This can be solved in various ways. If there is only the Ramond-Ramond $3$-form flux, the superfield $F^{KI}$ is symmetric, then we must have that 
\begin{align}
    \frac14 \chi^J(\sigma_{jk})_J{}^I H_{ijk}=0.
\end{align}
Using the special sigma matrices basis discussed in Appendix \ref{su3structure} and the complex decomposition of the Kalb-Ramond field-strength $H_{ijk}=(H_{{\mathsf i}{\mathsf j}{\mathsf k}},H_{\bar{\mathsf i}\bar{\mathsf j}\bar{\mathsf k}},H_{\bar{\mathsf i}{\mathsf j}{\mathsf k}},H_{{\mathsf i}\bar{\mathsf j}\bar{\mathsf k}})$, the last equation can be solved with 
\begin{align}
    H_{{\mathsf i}{\mathsf j}{\mathsf k}}=H_{\bar{\mathsf i}\bar{\mathsf j}\bar{\mathsf k}}=
    \delta^{{\mathsf j}\bar{\mathsf i}}H_{\bar{\mathsf i}{\mathsf j}{\mathsf k}}=
    \delta^{{\mathsf i}\bar{\mathsf j}}H_{{\mathsf i}\bar{\mathsf j}\bar{\mathsf k}}=0.
\end{align}
It is important to observe that it is not possible to set the whole superfield $H_{ijk}=0$ since its 
$\theta^2$ terms contain the curvature. 

The first non-trivial implication of (\ref{Nphi}) is, using (\ref{NNPhi}), (\ref{NtildeNPhi})  and that $\nabla_{\beta J}\chi^I\Big|_{\theta=0}=0$ we obtain
\begin{align}
    \nabla_{\beta J}\left(\chi^I\nabla_{\alpha}\Phi +\chi^I\widetilde\nabla_{\alpha I}\Phi\right)\Big|_{\theta=0}=\frac18 \epsilon_{\beta\alpha}\chi^I\left(-\sigma^i_{IJ}\nabla_i\Phi+4F_{IJ}\right)\Big|_{\theta=0}=0.
\end{align}
This equation can be solved non-trivially if the $SU(3)$ singlet part of the Ramond-Ramond $3$-form vanishes
\begin{align}
    \chi^I \chi^K F_{IK}=0,
\end{align}  
and the $\mathbf 3$ is equal to the derivative of the dilaton. The $\mathbf 6$ representation of the RR field does not appear in the equation above. 

Computing $V_i$ with the conditions above, we find 
\begin{align}
    V_i= \frac12|\chi|^2\nabla_i \Phi -\chi^I \chi_J \left( (\s_i)_{IK} F^{KJ} +(\s_i)^{JK} F_{KI} \right).
\end{align}
If we use this information in equation (\ref{secondF}) using $\chi=\widetilde\chi$ we obtain
\begin{align}
    -\frac12 \chi_I F = \frac12 |\chi|^2\chi^K(\sigma_i)_{KI}\nabla_i\Phi -2|\chi|^2\chi^K F_{IK}.
\end{align}
Contracting this last expression with $\chi^I$, we find that $F=0$. It is straightforward to check that (\ref{cosmologicalConstant}) also vanishes. Therefore, the fact that the $SU(3)$ singlet of the RR $3$-form vanishes implies that the cosmological constant is zero, up to terms with $\Omega^2$ and $\widetilde\Omega^2$. This is a known result; having an $AdS_4$ compactification in Type IIB with only one normalizable spinor is impossible \cite{Lust:2004ig, Caviezel:2008ik}.

\section{Conclusions and prospects}
\label{conclusion}

In this work, we have proposed a formalism to study compactifications of Type II supergravity theories that preserve $N=1$ supersymmetry in four dimensions, working directly in a curved superspace formulation inspired by the pure spinor superstring. 

The central hypothesis is the factorization of the massless vertex operators into a product of four-dimensional superfields and internal ``superspace harmonic form.'' These superspace harmonic forms can be considered the generalization of the harmonic forms in a usual Calabi-Yau compactification. This factorization relied on redefining the covariant derivatives in terms of auxiliary bosonic spinor superfields $(\chi, \widetilde\chi)$, separating the four-dimensional derivatives from the mixed components. This decomposition shows how the familiar four-dimensional superspace covariant derivatives emerge naturally from the BRST differential.

Our superspace approach provides a geometric framework for constructing and analyzing supersymmetric compactifications while maintaining manifest supersymmetry and lower-dimensional covariance. It elegantly allows us to directly implement symmetry conditions on the higher-dimensional backgrounds. The factorized form of the massless fields is handy, as it connects directly to the standard four-dimensional superfield formalism widely used in model-building and phenomenology.
Looking ahead, there are many potential applications and extensions of this work. Our formalism can be applied to study specific flux compactifications of interest. Furthermore, while our focus has been on $N=1$ supersymmetry in four dimensions, the methods developed here may also prove helpful for other amounts of global supersymmetry and other space-time dimensions. 

During our analysis, we found no condition that fixes the Grassmann odd covariant derivatives of  
$(\chi,\widetilde\chi)$ besides $\nabla_\alpha\chi^I=0$. On purely dimensional grounds, it should have the general form 
\begin{align}
    \nabla_{\alpha I}\chi^J = \Omega_{\alpha K}H^{KJ}{}_I(\Phi,\chi,\widetilde\chi)+\widehat\Omega_{\alpha\widetilde K}{\widetilde H}^{\widetilde K J}{}_I(\Phi,\chi,\widetilde\chi),
\end{align}
where $(H,\widetilde H)$ are internal superfields depending only on $\Phi$, $\chi$ and $\widetilde\chi$. There are analogous expressions with other derivatives. It may be possible to find the explicit expressions by demanding consistency with (\ref{internalChi}) and (\ref{pureSpinorEqs}). Related to this is the problem of finding the minimal set of superspace equations necessary for the supersymmetry algebra to hold. 

Another crucial point is to add $D$-branes or other extended objects to the curved background. It is well-known that compact solutions with Ramond-Ramond fluxes cannot exist without source terms for extended objects \cite{Maldacena:2000mw, Ivanov:2000fg, Giddings:2001yu}. These source terms will appear in some superspace Bianchi identities and will change the solutions to these identities found in Section \ref{PureSugraReview}. Since the Berkovits-Howe constraints are found from the world-sheet action, the presence of extended objects may change the constraints themselves. This is still an open problem in pure spinor formalism, and we plan to address it in the future.

The most natural next step is to study vertex operators in the formalism described here. One of the primary motivations for studying flux compactifications is to generate masses for the moduli describing deformations of the compactification manifold. Such mass terms should appear in the equations of motion for the four-dimensional $N=1$ chiral multiplets. The ten-dimensional covariant derivative algebra, with the fluxes appearing in them, should provide such mass terms. A simple comparison with the ten-dimensional case indicates that the scalar four-dimensional superfield in (\ref{factorizedVertex}) is the pre-potential of a chiral field. It would be interesting to verify these statements.

A more ambitious goal is to find hidden world-sheet symmetries in the case of a Calabi-Yau compactification. It is a textbook fact that the RNS formalism possesses $N=2$ superconformal symmetry in the world-sheet in these compactifications. This powerful symmetry is responsible for many simplifications in perturbative calculations and non-perturbative effects. The observable consequences of this symmetry should not depend on the formalism used; therefore, they should manifest in some way in the pure spinor formalism. A possible way for this to happen is that the composite $b$ \cite{Berkovits:2005bt, Chandia:2021coc} ghost of the pure spinor formalism factories in a space-time plus compactification and that the compactification part forms a closed world-sheet current algebra with the compactification part of the BRST current.\footnote{This can be defined in the same way we defined the space-time covariant derivatives in the present work.} 

A formalism that combines a manifest supersymmetry formulation with the preservation of world-sheet supersymmetry was recently introduced in \cite{Berkovits:2021xwh}; it incorporates characteristics of all previous superstring formalisms. This new approach was studied in the curved background for the heterotic string case \cite{Berkovits:2022dbm} and the type II superstring case  \cite{Chandia:2023eel}. If it is possible to write the action for a curved background so that the world-sheet supersymmetry is manifest, the number of background couplings will be significantly smaller. If that happens, analyzing the existence of global supersymmetries will be much simpler. This a line of investigation that is worth pursuing.

{\bf Acknowledgments:} We thank William D. Linch, III, for valuable comments and suggestions. OC would like to thank the Kavli Institute for Theoretical Physics (KITP), where part of this work was done. This work was partially financed by FONDECYT Regular grants 1200342 and 1201550. This research was partly supported by grant NSF PHY-2309135 to the Kavli Institute for Theoretical Physics.

\appendix

\section{Ten dimensional $\g$ matrices}\label{identities}
A useful notation for spinorial indices is the following. Express the $\bf{16}$ spinorial representation of $SO(1,9)$ as representations of $SL(2,C)$ and vector representation $SU(4)$. Then, the superspace coordinate $\t^\a$ (here $\a=1, \dots, 16)$ of ten dimensions is expressed as $(\t^{\a I}, \t^{\dot\a}_I)$ (here $\a$ and $\dot\a$ run from $1, 2$, and $I=1, \dots, 4$). The non-zero components of the ten-dimensional gamma matrices, $(\g^a)_{\a\b},  (\widetilde\gamma^a)^{\a\b}$,  are given by (the ten-dimensional vector index is split in the four-dimensional vector index $a$ and the six-dimensional vector index $i$) 
\begin{align}
&(\g^a)_{{}_{\a I},{{}_{\dot\b}^J}}=(\s^a)_{\a\dot\b}\d_I^J,\quad (\g^i)_{{}_{\a I},{}_{\b J}} =  (\s^i)_{IJ}\e_{\a\b},\quad  (\g^i)_{{}_{\dot\a}^I,{}_{\dot\b}^J} = -(\s^i)^{IJ} \e_{\dot\a\dot\b} ,\cr
&(\widetilde\g^a)^{{}_{\a I},{}^J_{\dot\b}}=-(\widetilde\s^a)^{\dot\b\a} \d_J^I,\quad (\widetilde\g^i)^{{}_{\a I},{}_{\b J}} = (\s^i)^{IJ}\e^{\a\b},\quad (\widetilde\g^i)^{{}_{\dot\a}^I,{}_{\dot\b}^J} = -(\s^i)_{IJ} \e^{\dot\a\dot\b} .
\label{g10}
\end{align}
where $\s^a, \widetilde\s^a, \e$'s are defined in \cite{Wess:1992cp}, and $\s^i$'s are defined to satisfy the Dirac algebra
\begin{align}
    (\s_{i})_{IK}(\s_{j})^{KJ}+(\s_{j})_{IK}(\s_{i})^{KJ}=2\d_{ij}\d_I^J.
\end{align}
Non-zero products of two gamma matrices are given by 
\begin{align}
&(\g_{ab})_{{}_{\a I}}{}^{{}_{\b J}}=-(\s_{ab})_\a{}^\b \d_I^J,\quad (\g_{ab})_{{}_{\dot\a}^I}{}~^{{}_{\dot\b}^J} = (\s_{ab})^{\dot\b}{}_{\dot\a} \d^I_J ,\cr
&(\g_{ij})_{{}_{\a I}}{}^{{}_{\b J}}=(\s_{ij})_I{}^J \d_\a^{\b}, \quad  (\g_{ij})_{{}_{\dot\a}^I}{}~^{{}_{\dot\b}^J} = (\s_{ij})^I{}_J \d^{\dot\b}_{\dot\a} ,\cr
&(\g_{ai})_{{}_{\a I}}{}~^{{}_{\dot\b}^J} = -(\s_a)_{\a\dot\g}\e^{\dot\g\dot\b}(\bar\s_i)_{IJ},\quad (\g_{ai})_{{}_{\dot\a}^I}{}~^{{}_{\b J}} = -\e^{\b\g}(\s_a)_{\g\dot\a} (\s_i)^{IJ} .
\label{}
\end{align}
Non-zero products of three gammas 
\begin{align}
&(\g_{abc})^{{}_{\a I}~{}_{\dot\b}^J}=-i\e_{abcd}(\widetilde\s^d)^{\dot\b\a}\d^I_J,\quad (\g_{abi})^{{}_{\a I}~{}_{\b J}}=-\e^{\b\g}(\s_{ab})_\g{}^\a(\s_i)^{IJ} ,\cr
&(\g_{abi})^{{}_{\dot\a}^I~{}_{\dot\b}^J}=(\s_{ab})^{\dot\a}{}_{\dot\g}\e^{\dot\g\dot\b}(\bar\s)_{IJ},\quad (\g_{aij})^{{}_{\a I}~{}_{\dot\b}^J}=-(\widetilde\s_a)^{\dot\b\a}(\s_{ij})^I{}_J ,\cr
&(\g_{ijk})^{{}_{\a I}~{}_{\b J}} = \e^{\a\b}(\s_{ijk})^{IJ},\quad (\g_{ijk})^{{}_{\dot\a}^I~{}_{\dot\b}^J} = -\e^{\dot\a\dot\b} (\s_{ijk})_{IJ} .
\label{}
\end{align}
Some useful identities involving four and six-dimensional $\s$ matrices are
\begin{align}\label{IDS}
    &(\s^a)_{\a\dot\b}(\s_a)^{\dot\g\r}=-2\d_\a^\r \d_{\dot\b}^{\dot\g} ,\cr 
    &(\s_{ab})_\a{}^\b(\s^{ab})_\g{}^\r=4\d_\a^\b \d_\g^\r-8\d_\g^\b \d_\a^\r ,\cr
    &(\s_{ab})_\a{}^\b(\s^{ab})^{\dot\g}{}_{\dot\r}=0,\cr
    &(\s^i)^{IJ}(\s^i)_{KL}=-2(\d^I_K\d^J_L-\d^I_L\d^J_K),\cr
    &(\s^i)_{IJ}(\s^i)_{KL}=-2\e_{IJKL},\cr 
    &(\s^{ij})^I{}_J(\s^{ij})^K{}_L=2\d^I_J\d^K_L-8\d^I_L\d^K_J .
\end{align}

\section{The calculation of $[\nabla_\alpha,S_{\beta J}\nabla_{\dot\beta}{}^J +\widetilde S_{\beta\widetilde J}\widetilde\nabla_{\dot\beta}{}^{\widetilde J}+S_{\dot\beta}{}^{J}\nabla_{\beta J} +\widetilde S_{\dot\beta}{}^{\widetilde J}\widetilde\nabla_{\beta\widetilde J}]$ and other identities}
\label{NablaSComputation}
In this appendix we include intermediate steps in the computation of $[\nabla_\alpha,S_{\beta J}\nabla_{\dot\beta}{}^J +\widetilde S_{\beta\widetilde J}\widetilde\nabla_{\dot\beta}{}^{\widetilde J}+S_{\dot\beta}{}^{J}\nabla_{\beta J} +\widetilde S_{\dot\beta}{}^{\widetilde J}\widetilde\nabla_{\beta\widetilde J}]$. We begin with the following identities
\begin{align}
   &\N_\a\O_{\b J}=\frac14\e_{\a\b}(\chi\s_i)_J\N_i\Phi+(\N_{\b J}\chi^I)\O_{\a I}+(\N_{\b J}\widetilde\chi^{\widetilde I})\Oh_{\a\widetilde I},\\ 
   &\N_\a\Oh_{\b\widetilde J}=\frac14\e_{\a\b}(\widetilde\chi\s_i)_{\widetilde J}\N_i\Phi+(\N_{\b\widetilde J}\chi^I)\O_{\a I}+(\N_{\b\widetilde J}\widetilde\chi^{\widetilde I})\Oh_{\a\widetilde I},\\ 
   &\N_\a\O_{\dot\b}{}^J=(\N_{\dot\b}{}^J\chi^I)\O_{\a I}+(\N_{\dot\b}{}^J\widetilde\chi^{\widetilde I})\Oh_{\a\widetilde I},\\
   &\N_\a\Oh_{\dot\b}{}^{\widetilde J}=(\N_{\dot\b}{}^{\widetilde J}\chi^I)\O_{\a I}+(\N_{\dot\b}{}^{\widetilde J}\widetilde\chi^{\widetilde I})\Oh_{\a\widetilde I}.
\end{align}
Using these results we obtain
\begin{align}
    [\N_\a,S_{\b J}\N_{\dot\b}{}^J]&=\e_{\a\b}\left(|\chi|^2(\chi\s_i)_J\N_i\Phi-\frac14(\chi\s_i)_J\widetilde\chi^{\widetilde I}\widetilde\chi_{\widetilde K}(\s_j\s_i)_{\widetilde I}{}^{\widetilde K}\N_j\Phi \right.\cr
    &+\left.\frac1{16}(\chi\s_{ij})_J\widetilde\chi^{\widetilde I}\widetilde\chi^{\widetilde K}(\s_{ij}\s_k)_{\widetilde I\widetilde K} \N_k\Phi\right)\N_{\dot\b}{}^J\cr 
    &+(1-x)S_{\b J}\widetilde\chi^I\left( \N_{\a\widetilde I}\O_{\dot\b}{}^J S +\N_{\dot\b}{}^J\Oh_{\a\widetilde I} \widetilde S\right)+\cdots,
\end{align}
\begin{align}
    [\N_\a,\widetilde S_{\b\widetilde J}\N_{\dot\b}{}^{\widetilde J}]&=\e_{\a\b}\left(|\widetilde\chi|^2(\widetilde\chi\s_i)_J\N_i\Phi-\frac14(\widetilde\chi\s_i)_{\widetilde J}\chi^{I}\chi_{K}(\s_j\s_i)_{I}{}^{K}\N_j\Phi \right. \cr
    &+\left.\frac1{16}(\widetilde\chi\s_{ij})_{\widetilde J}\chi^{I}\chi^{K}(\s_{ij}\s_k)_{IK} \N_k\Phi\right)\N_{\dot\b}{}^{\widetilde J} \cr 
    &+(1-x)\widetilde S_{\b\widetilde J}\chi^I\left(\N_{\dot\b}{}^{\widetilde J}\O_{\a I} S + \N_{\a I}\Oh_{\dot\b}{}^{\widetilde J}\widetilde S\right)+\cdots ,
\end{align}
\begin{align}
    [\N_\a,S_{\dot\b}{}^J\N_{\b J}]&=\frac14(1-x)S_{\dot\b}{}^J\chi^I\e_{\a\b}(\s_i)_{IJ}\N_i\Phi S\cr 
    &+(1-x)S_{\dot\b}{}^J\widetilde\chi^{\widetilde I}\left(\N_{\a\widetilde I}\O_{\b J}S+\N_{\b J}\Oh_{\a\widetilde I}\widetilde S\right)+ \cdots ,
\end{align}
\begin{align}
    [\N_\a,\widetilde S_{\dot\b}{}^{\widetilde J}\N_{\b\widetilde J}]&=\frac14(1-x)\widetilde S_{\dot\b}{}^{\widetilde J}\widetilde\chi^{\widetilde I}\e_{\a\b}(\s_i)_{\widetilde I\widetilde J}\N_i\Phi\widetilde S \cr 
    &+(1-x)\widetilde S_{\dot\b}{}^{\widetilde J}\chi^{I}\left(\N_{\a I}\Oh_{\b\widetilde J}\widetilde S+\N_{\b\widetilde J}\O_{\a I} S\right)+\cdots .
\end{align}
The $\cdots$ in the commutators above include terms with $\Omega^2$, $\widetilde\Omega^2$ and $\Omega\widetilde\Omega$. We do not include them explicitly since these terms need at least two Grassmann odd derivatives to have a non-vanishing contribution in the covariant $\theta$ expansion. 

Other useful identifies used in some computations are 
\begin{align}\label{NtildeNPhi}
    &(\g^{\underline a}P\g_{\underline a})_{\a I\b \tilde J} =2\e_{\a\b} P_{\dot\g I}{}^{\dot\g}{}_{\tilde J}+(\s^i)_{IK}(\s^i)_{\tilde J\tilde L} P_\a{}^K{}_\b{}^{\tilde L}  
    =-8\N_{\a I}\Oh_{\b\tilde J}=8\N_{\b\tilde J}\O_{\a I},
\end{align}
\begin{align}
    &(\g^{\underline a}P\g_{\underline a})_{\a I\dot\b{}^{\tilde J}} = 2P_{\dot\b I\a}{}^{\tilde J}-(\s^i)_{IK}(\s^i)^{\tilde J\tilde L} P_{\a}{}^K{}_{\dot\b\tilde L} 
    =-8\N_{\a I}\Oh_{\dot\b}{}^{\tilde J}=8\N_{\dot\b}{}^{\tilde J}\O_{\a I},
\end{align}
\begin{align}
    &(\g^{\underline a}P\g_{\underline a})_{\dot\a{}^I\b\tilde J}=2P_\b{}^I{}_{\dot\a\tilde J}-(\s^i)^{IK}(\s^i)_{\tilde J\tilde L}P_{\dot\a K\b}{}^{\tilde L} 
    =-8\N_{\dot\a}{}^I\Oh_{\b\tilde J}=8\N_{\b\tilde J}\O_{\dot\a}{}^I,
\end{align}
\begin{align}
    &(\g^{\underline a}P\g_{\underline a})_{\dot\a{}^I\dot\b{}^{\tilde J}}=2\e_{\dot\a\dot\b}P_\g{}^{I\g\tilde J}+(\s^i)^{IK}(\s^i)^{\tilde J\tilde L} P_{\dot\a K\dot\b\tilde L}
    =-8\N_{\dot\a}{}^I\Oh_{\dot\b}{}^{\tilde J}=8\N_{\dot\b}{}^{\tilde J}\O_{\dot\a}{}^I.
\end{align}

\section{Sigma matrices for $SU(3)$ Structure}
\label{su3structure}

The case where the bosonic part of the compactification manifold has only one normalizable spinor (in our work, this implies that $\widetilde\chi$ is proportional to $\chi$) is commonly said that the space has $SU(3)$ structure. This Appendix will describe a particular basis for the sigma matrices. 
We will use complex indices for the internal directions and an $\mathfrak{su}(3)$ decomposition of the $\mathfrak{so}(6)$ algebra acting on the tangent space indices of the internal superspace. The complex indices will range from $1$ to $3$ and we will use $\mathsf{k},\bar{\mathsf{k}},\mathsf{l},\bar{\mathsf{l}},\mathsf{m},\bar{\mathsf{m}}$ {\it etc}. and are raised and lowered with $(\delta^{\mathsf{k}\bar{\mathsf{l}}},\delta_{\mathsf{k}\bar{\mathsf{l}}})$. Index contraction can be written in different ways, as $a^ib_i=\frac12(\delta_{\mathsf{k}\bar{\mathsf{l}}}a^\mathsf{k} b^{\bar{\mathsf{l}}}+\delta_{\mathsf{k}\bar{\mathsf{l}}}b^\mathsf{k} a^{\bar{\mathsf{l}}})=\frac12(a^\mathsf{k}b^{\bar{\mathsf{k}}}+b^\mathsf{k}a^{\bar{\mathsf{k}}})$. The full $\mathfrak{so}(1,9)$ is then generated by $(M_{ab},M_{a\mathsf{k}},M_{a\bar{\mathsf{k}}},M_{\mathsf{k}\bar{\mathsf{l}}},M_{\mathsf{k}\mathsf{l}},M_{\bar{\mathsf{k}}\bar{\mathsf{l}}},M_{\bf 1})$. The internal Lorentz generators are $(M_{\mathsf{k}\bar{ \mathsf{l}}},M_{\mathsf{k}\mathsf{l}},M_{\bar{\mathsf{k}}\bar{\mathsf{l}}},M_{\bf 1})$, where $\delta^{\mathsf{k}\bar{\mathsf{l}}}M_{\mathsf{k}\bar{\mathsf{l}}}=0$.

In the case where $\widetilde\chi$ is proportional to $\chi$, an special basis can be chosen for the $\sigma$-matrices such that
\begin{align}
    (\sigma^{\bar{\mathsf{k}}})_{IJ}\chi^J=(\sigma^\mathsf{k})^{IJ}\chi_J=0.
\end{align}
We also have the following properties 
\begin{align}\label{SpinorProperties}
    &M^{\mathsf{k}\bar{\mathsf{l}}}\chi^I=M^{\mathsf{k}\bar{\mathsf{l}}}\chi_I=0,\quad M_{\bf 1}\chi^I=\frac12 \chi^I,\quad 
    M_{\bf 1}\chi_I=-\frac12 \chi_I,\cr 
    &M^{\mathsf{kl}}\chi^I=\frac12(\sigma^{\mathsf{kl}})_J{}^I\chi^J,\quad M^{\mathsf{kl}}\chi_I=0,\quad
    M^{\bar{\mathsf{k}}\bar{\mathsf{l}}}\chi^I=0,\quad M^{\bar{\mathsf{k}}\bar{\mathsf{l}}}\chi_I=\frac12(\sigma^{\bar{\mathsf{k}}\bar{\mathsf{l}}})_I{}^J\chi_J.
\end{align}

The anti-symmetric products of two sigma matrices are
\begin{align}
    (\sigma^{\mathsf{k}\bar{\mathsf{l}}})_I{}^J=\frac12\left([\sigma^{\mathsf{k}},\sigma^{\bar{\mathsf{l}}}]\right)_I{}^J,
    \quad (\sigma^{\mathsf{kl}})_I{}^J=\frac12\left([\sigma^{\mathsf{k}},\sigma^{\mathsf{l}}]\right)_I{}^J,\quad 
    (\sigma^{\bar{\mathsf{k}}\bar{\mathsf{l}}})_I{}^J=\frac12\left([\sigma^{\bar{\mathsf{k}}},\sigma^{\bar{\mathsf{l}}}]\right)_I{}^J.
\end{align}

The spinor realizations of the internal Lorentz generators are 
\begin{align}
    &\left(M_{\bf 1}\right)_I{}^J=\frac16 \delta_{\mathsf{k}\bar{\mathsf{l}}}\left(\sigma^{\mathsf{k}\bar{\mathsf{l}}}\right)_I{}^J\equiv \frac12 (\sigma_{\bf 1})_I{}^J,\quad
    \left(M^{\mathsf{k}\bar{\mathsf{l}}}\right)_I{}^J= \frac12\left( (\sigma^{\mathsf{k}\bar{\mathsf{l}}})_I{}^J-\delta^{\mathsf{k}\bar{\mathsf{l}}}\left(\sigma_{\bf 1}\right)_I{}^J\right),\\
    &\left(M^{\mathsf{kl}}\right)_I{}^J=\frac12 (\sigma^{\mathsf{kl}})_I{}^J,\quad 
    \left(M^{\bar{\mathsf{k}}\bar{\mathsf{l}}}\right)_I{}^J=\frac12 (\sigma^{\bar{\mathsf{k}}\bar{\mathsf{l}}})_I{}^J,
\end{align}
so one can see that equations (\ref{SpinorProperties}) are satisfied.

The anti-symmetric product of three sigma matrices with holomorphic and anti-holomorphic indices will be defined as 
\begin{align}
    &(\sigma^{\mathsf{klm}})_{IJ}=\epsilon^{\mathsf{klm}}(\sigma^1\sigma^2\sigma^3)\equiv\epsilon^{\mathsf{klm}}(\sigma_{\Omega})_{IJ},\\
    &(\sigma^{\overline{\mathsf{klm}}})^{IJ}=\epsilon^{\overline{\mathsf{klm}}}(\sigma^{\bar 1}\sigma^{\bar 2}\sigma^{\bar 3})\equiv\epsilon^{\overline{\mathsf{klm}}}(\sigma_{\bar\Omega})^{IJ}.
\end{align}
Note that these matrices are symmetric in $IJ$ and that 
\begin{align}
    (\sigma_{\Omega})^{IJ}=\left((\sigma_{\bar\Omega})_{IJ}\right)^\dagger,\quad 
    (\sigma_{\bar\Omega})^{IJ}=\left((\sigma_{\Omega})_{IJ}\right)^\dagger
\end{align}
Another property is that 
\begin{align}
    \chi_I = f \left(\sigma_\Omega\chi\right)_I,\quad \chi^I= \bar f \left(\sigma_{\bar\Omega}\chi\right)^I,
\end{align}
for some internal superfield $f$. These also imply that 
\begin{align}
    \left(\sigma^{\bar{\mathsf{k}}}\chi\right)^I=\frac{f}{2}\epsilon^{\overline{\mathsf{klm}}} 
    \left(\sigma_{\overline{\mathsf{lm}}}\chi\right)^I,\quad 
    \left(\sigma^{\mathsf{k}}\chi\right)_I=\frac{\bar f}{2}\epsilon^{\mathsf{klm}} 
    \left(\sigma_{\mathsf{lm}}\chi\right)_I.
\end{align}

The remaining linearly independent products of three sigma matrices are $(\{\sigma^{\mathsf{kl}},\sigma^{\bar{\mathsf{m}}}\})_{IJ}$ and $(\{\sigma^{\overline{\mathsf{kl}}},\sigma^{\mathsf{m}}\})_{IJ}$. They can be decomposed into irreducible $\mathfrak{su}(3)$ representations using $\delta_{\mathsf{k}\bar{\mathsf{l}}}$. We will define them as 
\begin{align}
    &(\sigma^{\mathsf{k}}_{\mathbf 3})_{IJ}=\frac12\delta_{\mathsf{l}\bar{\mathsf{m}}}(\{\sigma^{\mathsf{kl}},\sigma^{\bar{\mathsf{m}}}\})_{IJ}, \quad 
    (\sigma^{\mathsf{kl}\bar{\mathsf{m}}}_{\mathbf 6})_{IJ}=(\{\sigma^{\mathsf{kl}},\sigma^{\bar{\mathsf{m}}}\})_{IJ}-\frac12 (\sigma^{[\mathsf{k}}_{\mathbf 3})_{IJ}\delta^{\mathsf{l}]\bar{\mathsf{m}}},\\
    &(\sigma^{\bar{\mathsf{k}}}_{\bar{\mathbf 3}})_{IJ}=\frac12\delta_{\mathsf{l}\bar{\mathsf{m}}}(\{\sigma^{\overline{\mathsf{km}}},\sigma^{\mathsf{l}}\})_{IJ}, \quad 
    (\sigma^{\overline{\mathsf{kl}}\mathsf{m}}_{\mathbf 6})_{IJ}=(\{\sigma^{\overline{\mathsf{kl}}},\sigma^{\mathsf{m}}\})_{IJ}+\frac12 \delta^{\mathsf{m}[\bar{\mathsf{l}}}(\sigma^{\bar{\mathsf{k}}]}_{\bar{\mathbf 3}})_{IJ}.
\end{align}
$\sigma^{\mathsf{k}}_\mathbf{3}$ and $\sigma^{\bar{\mathsf{k}}}_\mathbf{3}$ can also be written in terms of $\{\sigma^{\mathsf{k}},\sigma_\mathbf{1}\}$ and $\{\sigma^{\bar{\mathsf{k}}},\sigma_\mathbf{1}\}$.
Note that 
\begin{align}
    (\chi\sigma^{\mathsf{k}}_{\mathbf 3}\chi)=(\chi\sigma^{\mathsf{kl}\bar{\mathsf{m}}}_{\mathbf 6}\chi)=
    (\chi\sigma^{\bar{\mathsf{k}}}_{\bar{\mathbf 3}}\chi)=
    (\chi\sigma^{\overline{\mathsf{kl}}\mathsf{m}}_{\mathbf 6}\chi)=0,
\end{align}
so the only non-vanishing contraction with two $\chi^I$ is $(\chi\sigma_\Omega\chi)$. Similarly, the only non-vanishing contraction with $(\chi^I,\chi_J)$ is with powers of $\sigma_\mathbf{1}$. 

{
\bibliographystyle{abe}
\bibliography{mybib}
}
\end{document}